\documentclass[11pt,oneside,letterpaper]{article}
\pdfoutput=1
\usepackage[section]{placeins}
\usepackage{amssymb}
\usepackage{amsmath}
\usepackage[dvips]{graphicx}
\usepackage{setspace}
\usepackage{amsfonts}
\usepackage{fancyhdr}
\usepackage{xcolor}
\usepackage{caption}
\usepackage{graphicx}
\usepackage[permil]{overpic}
\usepackage{rotating}
\usepackage{comment}
\usepackage{color}
\usepackage{cite}
\usepackage{braket}
\definecolor{darkgreen}{rgb}{0,0.5,0}
\definecolor{darkblue}{rgb}{0,0,0.6}
\definecolor{purple}{rgb}{0.4,.2,0.7}
\newcommand{\p}{\partial}

\newcommand{\be}{\begin{equation}}
\newcommand{\ee}{\end{equation}}

\usepackage{subcaption}

\usepackage[colorlinks=true,citecolor=darkgreen,linkcolor=black,urlcolor=purple]{hyperref}

\usepackage{pdfsync}

\makeatletter
\newcommand*{\defeq}{\mathrel{\rlap{%
                     \raisebox{0.3ex}{$\m@th\cdot$}}%
                     \raisebox{-0.3ex}{$\m@th\cdot$}}%
                     =} 
\makeatother

\DeclareMathOperator{\Tr}{Tr}
\def\be{\begin{eqnarray}}
\def\ee{\end{eqnarray}}

\newcommand{\la}{\langle}

\newcommand{\tr}{\textrm{Tr}\,}

\newcommand{\bea}{\begin{eqnarray}}
\newcommand{\eea}{\end{eqnarray}}
\def\ben{\begin{equation}}
\def\een{\end{equation}}

    \let\p=\phi \let\r=v

\let\la=\label

\def\be{\begin{equation}}
\def\ee{\end{equation}}
\def\ba{\begin{array}}
\def\ea{\end{array}}

\def\ba#1\ea{\begin{align}#1\end{align}}
\def\bs#1\es{\begin{split}#1\end{split}}

\renewcommand{\p}{\partial}

\newcommand{\Ssemi}{S_{\rm semi \text{-} cl}}
\newcommand{\GN}{G_N}

\allowdisplaybreaks  

\numberwithin{equation}{section}

\def\nref#1{(\ref{#1})}
\def \la {\label}   
\def \be {\begin{equation}}
\def \ee {\end{equation}}
	
\def \JM#1 {{\color{blue}  JM: #1 }}
\usepackage{framed}

\begin{document}

\begin{center}

~
\vskip5mm

{\LARGE  {
The entropy of Hawking radiation  \\
\ \\
}}

\vskip10mm

Ahmed Almheiri,$^1$\ \ Thomas Hartman,$^{2}$\ \ Juan Maldacena,$^{1}$\\  Edgar Shaghoulian,$^{2}$\ \  and Amirhossein Tajdini$^{2}$

\vskip5mm

{\it $^1$ Institute for Advanced Study, Princeton, New Jersey, USA } \\
\vskip5mm
{\it $^2$ Department of Physics, Cornell University, Ithaca, New York, USA
} 

\vskip5mm

\end{center}

\vspace{4mm}

\begin{abstract}
\noindent
In this review, we describe recent progress on the black hole information problem that involves a new understanding of how to calculate the entropy of Hawking radiation. We show how the method for computing gravitational fine-grained entropy, developed over the past 15 years, can be extended to capture the entropy of Hawking radiation. This technique reveals large corrections needed for the entropy to be consistent with unitary black hole evaporation.

 \end{abstract}

\pagebreak
\pagestyle{plain}

\setcounter{tocdepth}{2}
{}
\vfill
\tableofcontents

\newpage

  \begin{spacing}{1.15}

\section{Introduction}


In this review we discuss some recent progress on aspects of the black hole information paradox. 

Before delving into it let us discuss a big picture motivation. 
One of the main motivations to study quantum gravity is to understand the 
earliest moments of the universe, where we expect that quantum effects are dominant. 
In the search for this theory, it is better to consider simpler problems. A simpler problem involves black holes. They also contain a singularity in  their interior. It is an anisotropic big crunch singularity, but it is also a situation where quantum gravity is necessary, making it difficult to analyze. Black holes, however, afford us the opportunity to study them as seen from the outside. This is simpler because far from the black hole we can neglect the effects of gravity and we can imagine asking sharp questions probing the black hole from far away. One of these questions will be the subject of this review. We hope that, by studying these questions, we will eventually understand the black hole singularity and learn some lessons for the big bang, but we will not do that here. 

Studies of black holes in the '70s showed that black holes behave as thermal objects. They have a temperature that leads to Hawking radiation. They also have an entropy given by the area of the horizon. This suggested that, from the point of view of the outside, they could be viewed as an ordinary quantum system.   
Hawking objected to this idea through what we now know as the ``Hawking information paradox." He argued that a black hole would destroy quantum information, and that the von Neumann entropy of the universe would increase by the process of black hole formation and evaporation. 
 Results from the '90s using string theory, a theory of quantum gravity, provided some precise ways to study this problem for very specific gravity theories. These results strongly suggest that information does indeed come out. However, the current understanding requires certain dualities to quantum systems  where the geometry of spacetime is not manifest. 
 
 During the past 15 years, a better understanding of the von Neumann entropy for gravitational systems was developed.   The computation of the entropy involves also an area of a surface, but the surface is not the horizon. It is a surface that minimizes the generalized entropy. This formula is almost as simple as the Bekenstein formula for black hole entropy \cite{Bekenstein:1972tm,Bekenstein:1973ur}. 
  More recently, this formula was applied to the black hole information problem, giving a new way to compute the entropy of Hawking radiation \cite{Penington:2019npb,Almheiri:2019psf}. The final result differs from Hawking's result and is consistent with unitary evolution.

The first version of the fine-grained entropy formula was discovered by Ryu and Takayanagi \cite{Ryu:2006bv}. It was subsequently refined and generalized by a number of authors \cite{Hubeny:2007xt,Lewkowycz:2013nqa,Barrella:2013wja,Faulkner:2013ana,Engelhardt:2014gca,Almheiri:2019psf,Penington:2019npb,Almheiri:2019hni}. Originally, the Ryu-Takayanagi formula was proposed to calculate holographic entanglement entropy in anti-de Sitter spacetime, but the present understanding of the formula is much more general. It requires neither holography, nor entanglement, nor anti-de Sitter spacetime. Rather it is a general formula for the fine-grained entropy of quantum systems coupled to gravity.

  Our objective is to review these results for people with minimal background in this problem. We will not follow a historical route but rather try to go directly to the final formulas and explain how to use them. For that reason,  we will not discuss many related ideas that served as motivation, or that are also very useful for the general study of quantum aspects of black holes. A sampling of related work includes  \cite{Zhao:2019nxk, Akers:2019nfi, Rozali:2019day, Chen:2019uhq, Bousso:2019ykv, Almheiri:2019psy, Chen:2019iro, Laddha:2020kvp, Mousatov:2020ics, Kim:2020cds, Saraswat:2020zzf, Chen:2020wiq, Marolf:2020xie, Verlinde:2020upt, Giddings:2020yes, Liu:2020gnp, Pollack:2020gfa, Balasubramanian:2020hfs, Gautason:2020tmk,Anegawa:2020ezn, Hollowood:2020cou, Krishnan:2020oun, Banks:2020zrt, Geng:2020qvw}.   

Many details and caveats will necessarily be swept under the rug, although we will discuss some potential technical issues in the discussion section. We believe the caveats are only technical and unlikely to change the basic picture.


\section{Preliminaries}

\subsection{Black hole thermodynamics}
When an object is dropped into a black hole, the black hole responds dynamically. The event horizon ripples briefly, and then quickly settles down to a new equilibrium at a larger radius. It was noticed in the 1970s that the resulting small changes in the black hole geometry are constrained by equations  closely parallel to the laws of thermodynamics \cite{Christodoulou:1970wf,Christodoulou:1972kt,Hawking:1971tu,Bekenstein:1972tm,Bekenstein:1973ur,carter1972rigidity,Bardeen:1973gs,Hawking:1974rv,Hawking:1974sw}. The equation governing the response of a rotating black hole is \cite{Bardeen:1973gs}
\be
\frac{\kappa}{8\pi \GN} d\left( \mbox{Area} \right)  = dM  - \Omega dJ \ ,
\ee
where $\kappa$ is its surface gravity\footnote{Unfortunately, the name ``surface gravity'' is a bit misleading since the proper acceleration of an observer hovering at the horizon is infinite. $\kappa$ is related to the force on a massless (unphysical) string at infinity, see e.g. \cite{Wald:1984rg}.}, $M$ is its mass, $J$ is its angular momentum, and $\Omega$ is the rotational velocity of the horizon. The area refers to the area of the event horizon, and $\GN$ is Newton's constant.
If we postulate that the black hole has temperature $T \propto \kappa$, and entropy $S_{\rm BH} \propto \mbox{Area}$, then this looks identical to the first law of thermodynamics in the form
\be\label{firstlaw}
TdS_{\rm BH} = dM - \Omega dJ \ .
\ee
In addition, the area of the horizon always increases in the classical theory \cite{Hawking:1971tu}, suggesting a connection to the second law of thermodynamics. 
This is just a rewriting of the Einstein equations in suggestive notation, and initially, there was little reason to believe that it had anything to do with `real' thermodynamics. In classical general relativity, black holes have neither a temperature nor any significant entropy.  This changed with Hawking's discovery that,  when general relativity is coupled to quantum field theory, black holes have a temperature \cite{Hawking:1974sw}
\be
T = \frac{\hbar \kappa}{2\pi } \ .
\ee
(We set $c=k_B = 1$.) This formula for the temperature fixes the proportionality constant in $S_{\rm BH} \propto \mbox{Area}$. The total entropy of a black hole and its environment also has a contribution from the quantum fields outside the horizon. This suggests that the total or `generalized' entropy of a black hole is \cite{Bekenstein:1973ur}
\be\label{sgen}
S_{\rm gen} = \frac{\mbox{Area of horizon}}{4 \hbar \GN} + S_{\rm outside}  \ , \ee
where $S_{\rm outside}$ denotes the entropy of  matter  as well as gravitons outside the black hole, as it appears in the semiclassical description. It also includes a vacuum contribution from the quantum fields \cite{Bombelli:1986rw}.\footnote{  The quantum  contribution by itself has an ultraviolet divergence from the short distance entanglement of quantum fields across the horizon. This piece is proportional to the area, $A/\epsilon_{uv}^2$.  However, matter loops also lead to an infinite renormalization of Newton's constant, $1/(4 G_N) \to { 1 \over 4 G_N} - { 1 \over \epsilon^2_{uv}}$.  Then these two effects cancel each other so that $S_{\rm gen}$ is finite.   As usual in effective theories, these formally ``infinite'' quantities are actually subleading when we remember that we should take a small cutoff but not too small, $l_p \ll \epsilon_{uv} \ll r_s$. } 
The generalized entropy, including this quantum term, is also found to obey the second law of thermodynamics \cite{Wall:2011hj}, 
\be
\Delta S_{\rm gen} \geq 0 \ ,
\ee
giving further evidence that it is really an entropy. This result is stronger than the classical area theorem because it also covers phenomena like Hawking radiation, when the area decreases but the generalized entropy increases due to the entropy of Hawking radiation. 

The area is measured in Planck units, $l_p^2 = \hbar \GN$, so if this entropy has an origin in statistical mechanics then a black hole must have an enormous number of degrees of freedom. For example, the black hole at the center of the Milky Way,   Sagittarius A*,  has
\be
S  \approx {10^{85}} \ .  
\ee
Even for a black hole the size of a proton, $S \approx 10^{40}$.
In classical general relativity, according to the no-hair theorem, there is just one black hole with mass $M$ and angular momentum $J$, so the statistical entropy of a black hole is naively zero. 
 Including quantum fields helps, but has not led to a successful accounting of the entropy. 
 Finding explicitly the states giving rise to the entropy is an interesting problem, which we will not discuss in this review.

\subsection{Hawking radiation}

The metric of a Schwarzschild black hole is
\be\label{schw}
ds^2 = -\left( 1 - \frac{r_s}{r} \right) dt^2 + \frac{dr^2}{1 - \frac{r_s}{r}} + r^2 d\Omega_2^2 \ .
\ee
The Schwarzschild radius  $r_s = 2\GN M$ sets the size of the black hole. We will ignore the angular directions $d\Omega_2^2$ which do not play much of a role. To zoom in on the event horizon, we change coordinates, $r \to r_s(1 + \frac{\rho^2}{4r_s^2})$, $t \to 2 r_s \tau$, and expand for $\rho \ll r_s$. This gives the near-horizon metric
\be \la{RescMe}
ds^2 \approx  -\rho^2 d\tau^2 + d\rho^2 \ .
\ee
To this approximation, this is just flat Minkowski spacetime. To see this, define the new coordinates
\be\label{rindlerc}
x^0 = \rho \sinh \tau , \qquad x^1 = \rho \cosh \tau
\ee
in which 
\be \la{LocMin}
ds^2\approx  -\rho^2 d\tau^2 + d\rho^2  = -(dx^0)^2 + (dx^1)^2 \ .
\ee
Therefore according to a free-falling observer, the event horizon $r=r_s$ is not special. It is just like any other point in a smooth spacetime, and in particular, the geometry extends smoothly past the horizon into the black hole. This is a manifestation of the equivalence principle: free-falling observers do not feel the effect of gravity.  Of course, an observer that crosses the horizon will not be able to send signals to the outside\footnote{We can say that the interior lies behind a Black Shield (or Schwarz Schild in German).}.

\begin{figure}[t]
\begin{center}
\includegraphics[scale=1]{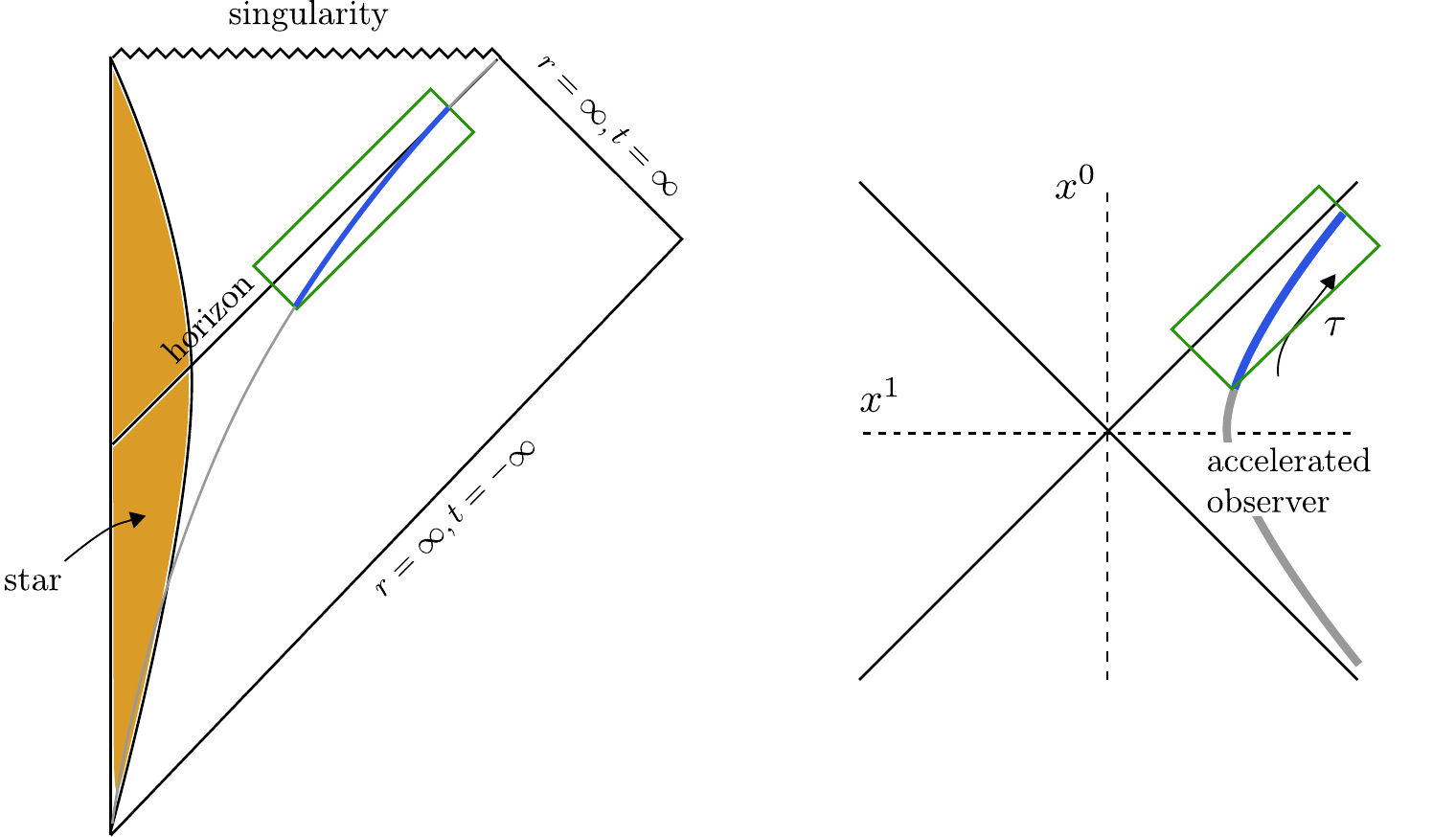}
\end{center}
\caption{\small Left: Penrose diagram of a black hole formed by gravitational collapse.  Right: Zoomed-in view of the flat near-horizon region, with the trajectory of a uniformly accelerated observer at $\rho = a^{-1}$.
\label{fig:schwarzschild}}
\end{figure}

The spacetime geometry of a Schwarzschild black hole that forms by gravitational collapse is illustrated in fig.~\ref{fig:schwarzschild}. An observer hovering near the event horizon at fixed $r$ is accelerating --- a rocket is required to avoid falling in. In the near-horizon coordinates \nref{LocMin}, an observer at fixed $\rho$ is following  the trajectory of a uniformly accelerated observer in Minkowski spacetime.

A surprising fact is that a uniformly accelerating observer in flat space detects thermal radiation. 
This is known as the Unruh effect \cite{Unruh:1976db}. There is a simple trick to obtain the temperature \cite{Bisognano:1976za}. The coordinate change \eqref{rindlerc} is very similar to the usual coordinate change from Cartesian coordinates to polar coordinates. It becomes identical if we perform the Wick rotation $\tau = i \theta$, $x^0 = i x^0_E$; then
\be
x^0_E = \rho \sin \theta , \quad x^1 = \rho \cos\theta \ .
\ee
The new coordinates $(x^0_E, x^1)$ or $(\rho, \theta)$ are simply Cartesian or polar coordinates on the Euclidean plane $\mathbb{R}^2$. 
In Euclidean space, an observer at constant $\rho$ moves in a circle of length $ 2\pi \rho$. 
Euclidean time evolution on a circle is related to the computation of thermodynamic quantities for the original physical system (we will return to this in section \ref{ss:gibbonshawking}). 
Namely, $\tr[e^{ - \beta H}] $ is the partition function at temperature $T=1/\beta$. $\beta$ is the length of the Euclidean time evolution and the trace is related to the fact that we are on a circle. This suggests that the temperature that an accelerated observer feels is 
\be \la{PropT}
T_{proper} = { 1 \over 2 \pi \rho } = { a \over 2 \pi } = { \hbar \over k_B c } { a \over 2 \pi } 
\ee 
where $a$ is the proper acceleration and we also restored all the units in the last formula. 
Though this argument seems a bit formal, one can check that a physical accelerating thermometer would actually record this temperature \cite{Unruh:1976db}. 

Now, this is the proper temperature felt by an observer very close to the horizon. Notice that it is infinite at $\rho=0$ and it decreases as we move away. This decrease in temperature is consistent with thermal equilibrium in the presence of a gravitational potential. In other words, for a spherically symmetric configuration, in thermal equilibrium, the temperature obeys the Tolman relation \cite{Tolman:1930zza} 
\be 
 T_{proper}(r) \sqrt{-g_{\tau\tau} (r) }= {\rm constant .}
\ee
This formula tracks the redshifting of photons as they climb a gravitational potential. It says that locations at a higher gravitational potential feel colder to a local observer. 
 Using   
 the  polar-like coordinates  \nref{LocMin} and \nref{PropT} we   indeed get a constant equal to $1/(2\pi)$.
 Since this formula is valid also in the full geometry \nref{schw}, we can then use it to find the temperature that an observer far from the black hole would feel. We simply need to undo the rescaling of time we did just above \nref{RescMe} and go to large $r$ where 
 $g_{tt} = -1$ to find the temperature 
 \be\label{thawking}
T 
= T_{proper} (r\gg r_s) = \frac{1}{4\pi r_s}  \ .
\ee
This is the Hawking temperature. It is the temperature measured by an observer that is more than a few Schwarzschild radii away from the black hole.

\subsection{The Euclidean black hole}\label{ss:gibbonshawking}

We will now expand a bit more on the connection between Euclidean time and thermodynamics. We will then use it to get another perspective on thermal aspects of black holes. Sometimes Euclidean time $t_E$ is called imaginary time and Lorentzian time $t$ is called real time because of the Wick rotation $t = i t_E$ mentioned above.  

There are different ways to see that imaginary-time periodicity is the same as  a temperature. In a thermal state, the partition function is
\be
Z = \tr [ e^{-\beta H} ]\ .
\ee
Any observable such as $\tr[ {\cal O}(t) {\cal O}(0) e^{-\beta H}]$ is periodic under $t \to t + i \beta$, using ${\cal O}(t) = e^{i H t}{\cal O}e^{-i H t}$ and the cyclic property of the trace. 

A more general argument in quantum field theory is to recast the trace as a path integral. Real-time evolution by $e^{-i H t}$ corresponds to a path integral on a Lorentzian spacetime, so imaginary-time evolution, $e^{-\beta H}$, is computed by a path integral on a Euclidean geometry. The geometry is evolved for imaginary time $\beta$, and the trace tells us to put the same boundary conditions at both ends and sum over them. A path integral on a strip of size $\beta$ with periodic boundary conditions at the ends is the same as a path integral on a cylinder. Therefore in quantum field theory $Z = \tr e^{-\beta H}$ is calculated by a path integral on a Euclidean cylinder with $\theta = \theta  + \beta$. Any observables that we calculate from this path integral will automatically be periodic in imaginary time.

\begin{figure}
\begin{center}
\includegraphics[scale=1]{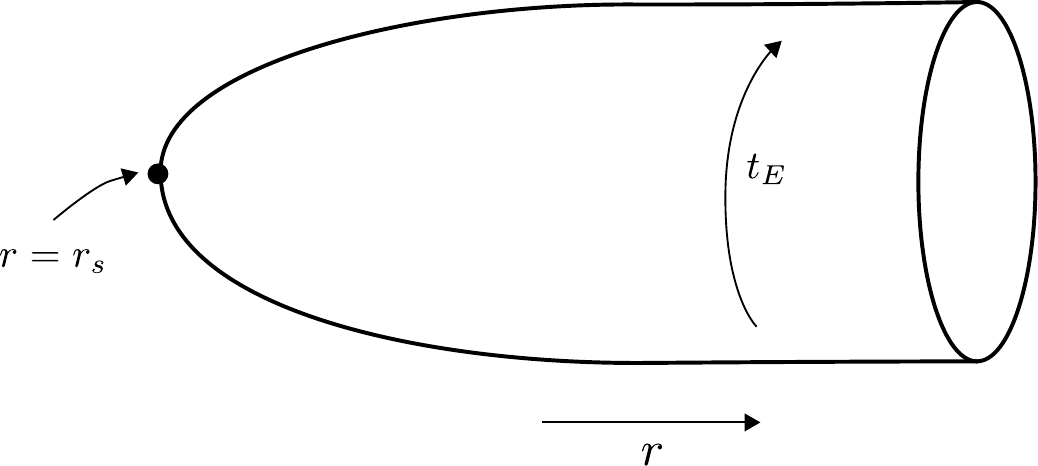}
\end{center}
\caption{\small  The Euclidean Schwarzschild black hole. The Euclidean time and radial directions have the geometry of a cigar, which is smooth at the tip, $r=r_s$. At each point we also have a sphere of radius $r$. \label{fig:cigar}}
\end{figure}

Similarly, in a black hole spacetime, the partition function at inverse temperature $\beta$ is calculated by a Euclidean path integral. The geometry is the Euclidean black hole, obtained from the Schwarzschild metric \eqref{schw} by setting $t = i t_E$,
\be \la{EuclBH}
ds^2_E = \left( 1 - \frac{r_s}{r} \right) dt_E^2 + \frac{dr^2}{1 - \frac{r_s}{r}} + r^2 d\Omega_2^2 \ , ~~~~~~~~~~ t_E = t_E + \beta \ .
\ee
In the Euclidean geometry, the radial coordinate is restricted to $r> r_s$, because we saw that $r-r_s$ is like the radial coordinate in polar coordinates, and $r=r_s$ is the origin   $-$ Euclidean black holes do not have an interior. In order to avoid a conical singularity at $r=r_s$ we need to adjust $\beta$ to 
\be 
\beta = 4 \pi r_s  \ .
\ee
 This geometry, sometimes called the `cigar,' is pictured in fig.~\ref{fig:cigar}. The tip of the cigar is the horizon. Far away, for $r \gg r_s$, there is a Euclidean time circle of circumference $\beta$, which is the inverse temperature as seen by an observer far away. Notice that in the gravitational problem we fix the length of the circle far away, but we let the equations determine the right radius in the rest of the geometry. 

The Euclidean path integral on this geometry is interpreted as the partition function,
\be
Z(\beta)  = \mbox{Path integral on the Euclidean black hole} \sim  e^{ - I_{\rm classical}} Z_{\rm quantum}  \ .
\ee
It has contributions from both gravity and quantum fields. The gravitational part comes from the Einstein action, $I$,  and is found by evaluating the action on the geometry \nref{EuclBH}. The quantum part is obtained by computing the partition function of the quantum fields on this geometry \nref{EuclBH}.
It is important that the geometry is completely smooth at $r=r_s$ and therefore the quantum contribution has no singularity there. 
 This is related to the fact that an observer falling into an evaporating  black hole sees nothing special at the horizon, as in the classical theory. 
 
 Then applying the standard thermodynamic formula to the result,
\be
S = (1 - \beta \p_\beta) \log Z(\beta)
\ee
gives the generalized entropy \eqref{sgen}. We will not give the derivation of this result but it uses that we are dealing with a solution of the equations of motion and that the non-trivial part of the variation can be concentrated near 
$r=r_s$  \cite{Gibbons:1977mu }.  

\subsection{Evaporating black holes}

\begin{figure}[p!]
\begin{center}
\begin{overpic}[grid=false,scale=1]{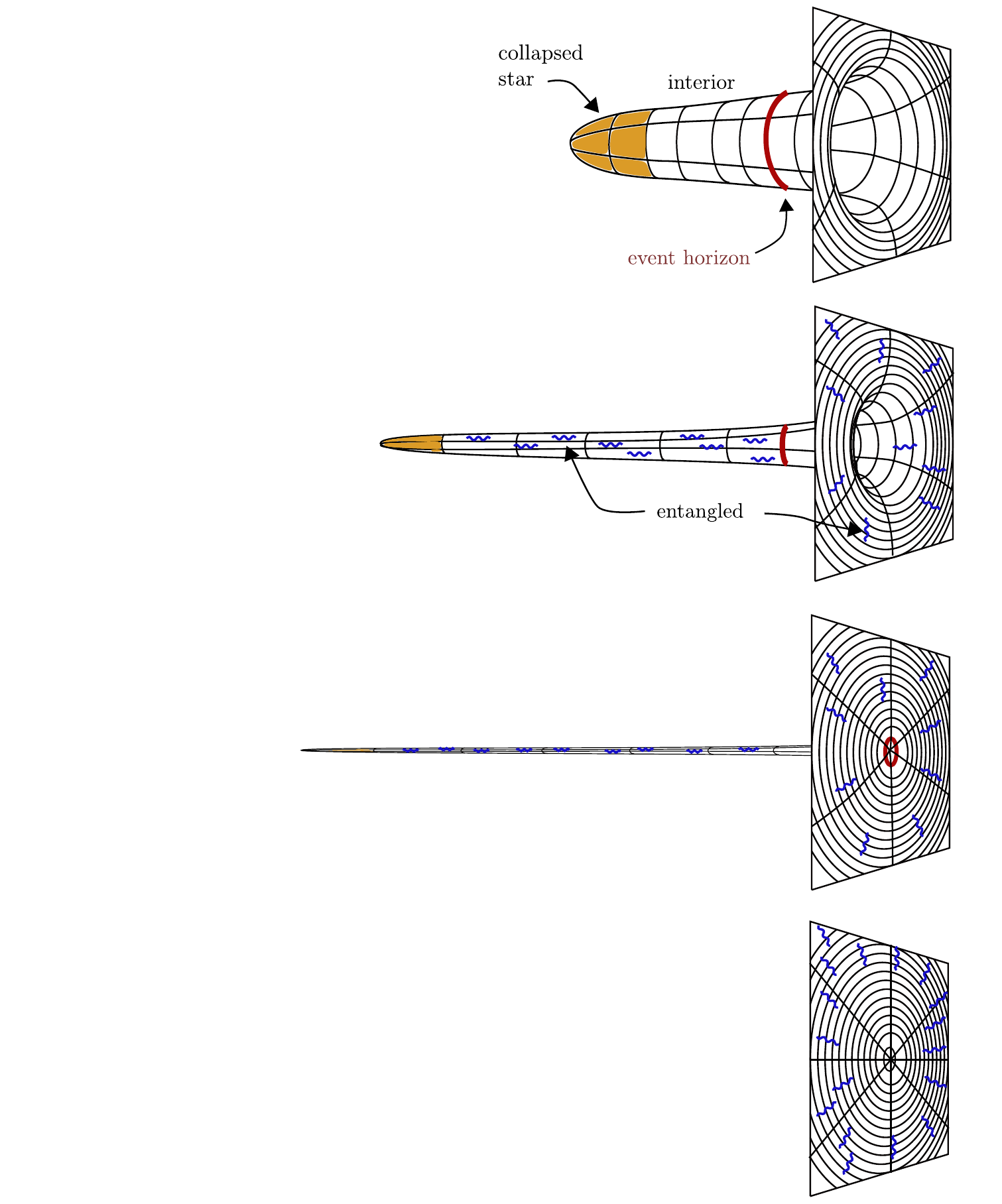}
\put(150,1050){ 
\parbox[t]{4in}{\centering \Large Stages of Black Hole Evaporation}
}
\put(50,950) {
\parbox[t]{2in}{$(a)$  After stellar collapse,
the outside of the black hole is nearly stationary, but on the inside, the geometry continues to elongate in one direction while pinching toward zero size in the angular directions.
}}
\put(50,720) {
\parbox[t]{4in}{
$(b)$
The Hawking process creates entangled pairs, one trapped behind the horizon
and the other escaping to infinity where it is observed as (approximate)
blackbody radiation. 
}}
\put(50,535) {
\parbox[t]{2in}{
The black hole slowly shrinks as its
mass is carried away by the radiation.
}}
\put(50,340) {
\parbox[t]{3.5in}{
$(c)$ Eventually the angular directions shrink to zero size. This is the singularity. The event horizon also shrinks to zero.
}}
\put(50,150) {
\parbox[t]{3.5in}{
$(d)$ At the end there is a smooth spacetime containing thermal Hawking radiation but no black hole.
}}
\end{overpic}
\end{center}
\caption{ \label{fig:evap-stages}}
\end{figure}

\begin{figure}
\begin{center}
\includegraphics[scale=1]{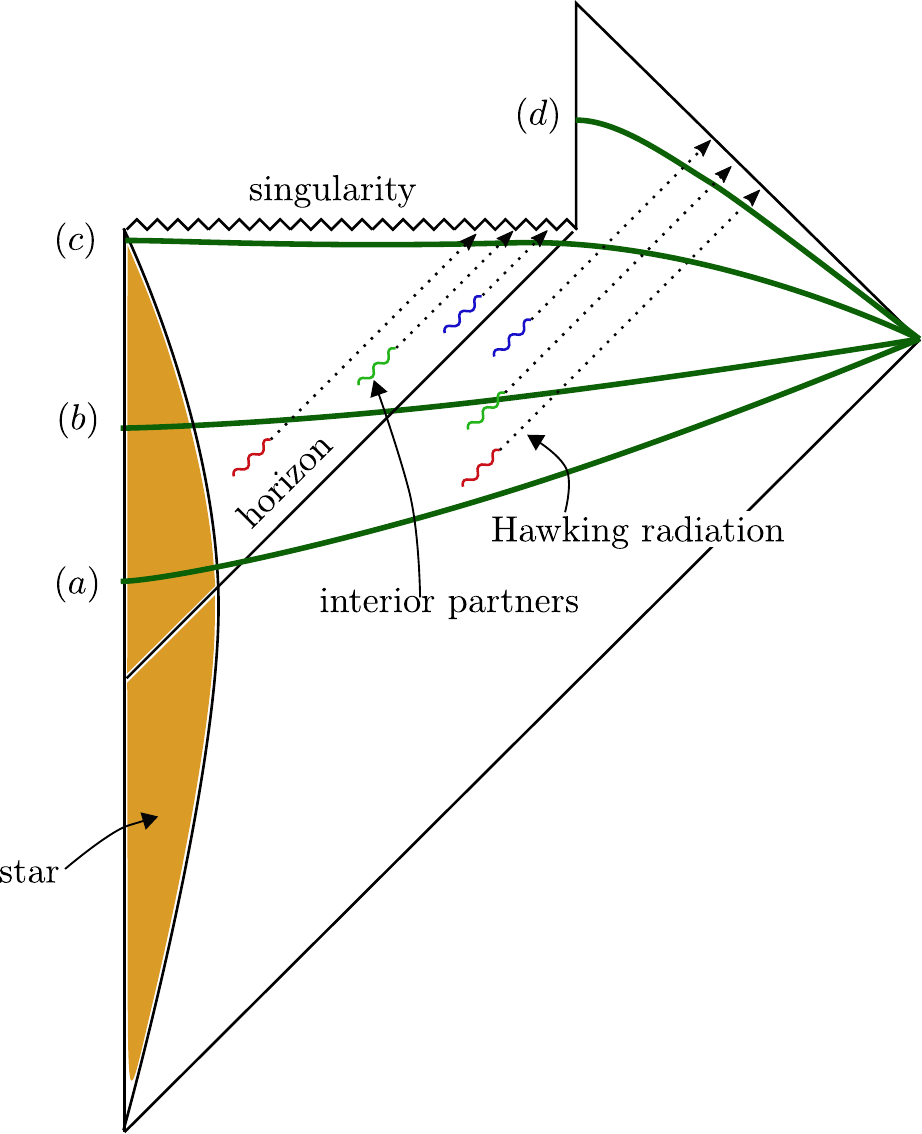}
\end{center}
\caption{Penrose diagram for the formation and evaporation of a black hole. Spatial slices $(a)$-$(d)$ correspond to the slices drawn in fig.~\ref{fig:evap-stages}. \label{fig:evap-penrose}}
\end{figure}

Hawking radiation carries energy away to infinity and therefore reduces the mass of the black hole. Eventually the black hole evaporates away completely --- a primordial black hole of mass $10^{12}\,$kg, produced in the early universe, would evaporate around now. The Hawking temperature of a solar mass black hole is $10^{-7}\,$K and its lifetime is $10^{64}$ years. 
The spacetime for this process is described in figures \ref{fig:evap-stages} and \ref{fig:evap-penrose}.

The Hawking process can be roughly interpreted as pair creation of entangled particles near the horizon, with one particle escaping to infinity and the other falling toward the singularity. This creation of entanglement is crucial to the story and we will discuss it in detail after introducing a few more concepts.

\section{The black hole as an ordinary quantum system} \label{central}


The results  we reviewed above suggest that the black hole can be regarded as an ordinary system, obeying the laws of thermodynamics. More precisely, as an object 
  described by a finite, but large, number of degrees of freedom that obey the ordinary laws of physics, which in turn imply the laws of thermodynamics.

In fact, this has been such an important idea in the development of the subject that we will call it the ``central dogma."   \begin{center}
{ \color{red} \underline{\bf%
 Central Dogma}}
\end{center}

{\center \begin{framed}   As seen from the outside, a black hole can be described in terms of a quantum system with  $  {\rm Area}/(4G_N ) $  degrees of freedom,   which evolves unitarily under time evolution. 
\end{framed}}

\begin{figure}[h]
\begin{center}
\includegraphics[scale=.3]{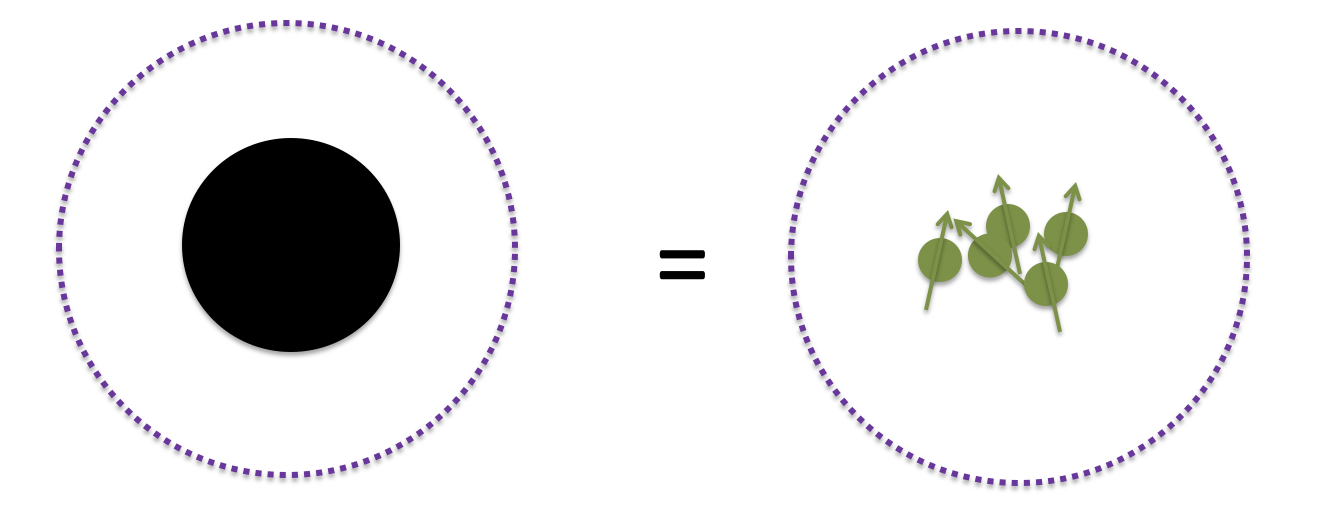}
\caption*{Using this hypothesis, the black hole and the whole spacetime around it, up to some surface denoted by the dotted circle, can be replaced by a quantum system. This quantum system interacts with the outside via a unitary Hamiltonian. }
\label{BHQuantumBdy}
\end{center}
\end{figure}

\begin{figure}[h]
\begin{center}
\includegraphics[scale=1]{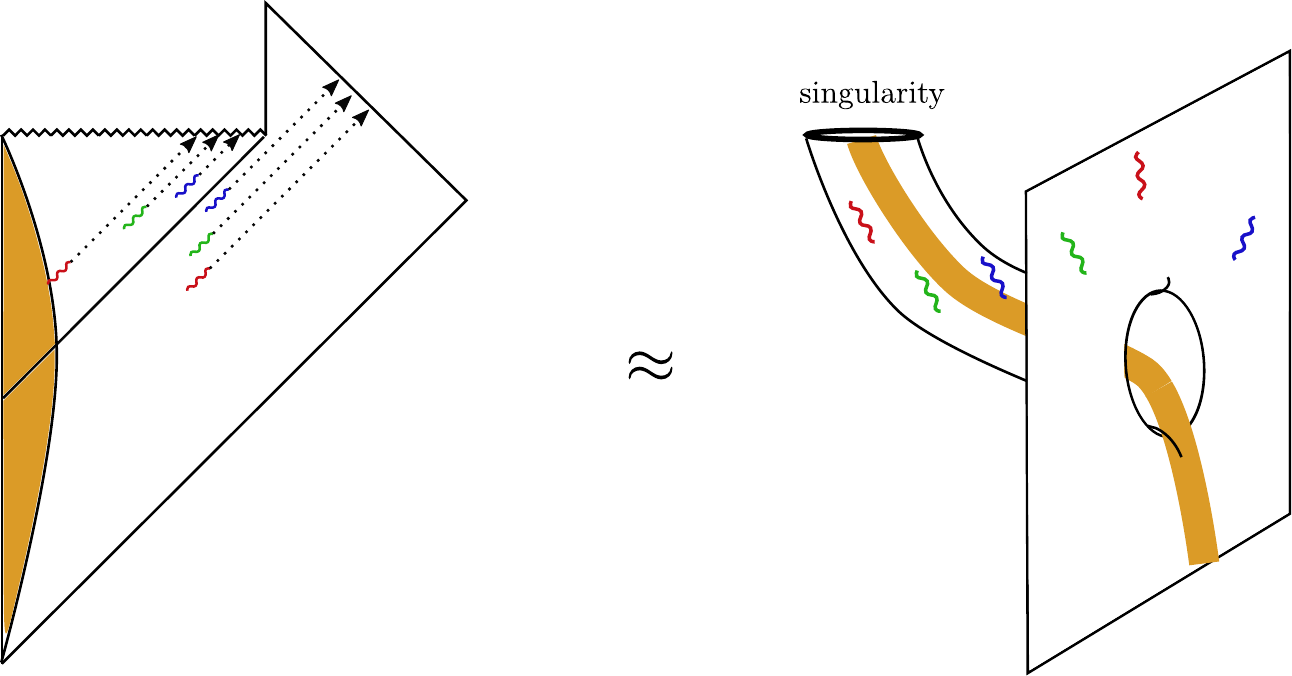} \\
(a) ~~~~~~~~~~~~~~~~~~~~~~~~~~~~~~~~~~~~~~~~~~~~~~~~~~~~~~~~~~~~~~~~~~~~~~~~~~~(b)
\caption{\small The skeptics' view: The diagram of an evaporating black hole is conceptually similar to one where we split off a baby universe, so that in the future we have two regions, the future region of the original universe, and the future of the interior, which is singular.}
\label{BabyUniverse}
\end{center}
\end{figure}

Let us make some remarks about this. 
\begin{itemize}
	\item 
	Notice that it is a statement about the black hole as seen from the outside. There is no statement about the black hole interior, yet.  
\item
The statement about the number of degrees of freedom is primarily a statement about the logarithm of the dimension  of the Hilbert space. We make no distinction between qubits, fermions or other degrees of freedom. What is important is that the Hilbert space has {\it finite} dimension.   
\item
The degrees of freedom that appear in this statement are not manifest in the gravity description. Some researchers have tried to see them as coming from the thermal atmosphere, by putting a cutoff, or ``brick wall'' at some Planck distance from the horizon \cite{tHooft:1984kcu,Thorne:1986iy}. Unfortunately, such ideas remain vague since such cutoffs are not manifest from the gravity point of view, which treats the horizon as a smooth surface. 
\item
Unitary evolution implies that we have a Hamiltonian that generates the time evolution. Again, this Hamiltonian is not manifest in the gravity description. The gravity description has a Hamiltonian constraint, which determines the bulk evolution. But this constraint is a property of the full spacetime, we do not know how to pull out a purely exterior part.  
In principle, the Hamiltonian could  be very general. But the fact that it gives rise to the gravity evolution constrains some properties. For example, it should be strongly interacting and generate a very chaotic evolution. 
\item
We said that the black hole evolves unitarily.   This is when we surround the black hole by a reflecting wall and we consider the full system inside this wall. However, if the black hole lives in an asymptotically flat geometry, it is convenient to draw an imaginary surface surrounding the black hole and call everything inside the ``quantum system'' that appears in the central dogma. This quantum system is then coupled   with the external degrees of freedom living outside this surface.  We usually think of the region outside the imaginary surface as a quantum system in a fixed spacetime, where we ignore large fluctuations of the background, though  we can still consider weakly interacting gravitons.  The full coupled evolution should be unitary. In other words, in this context, the gravity answers are compared to those of a quantum system that is coupled to the degrees of freedom far from the black hole at this imaginary cutoff surface. Here we are imagining that this surface is at a few Schwarzschild radii from the black hole. 
\item
Often people ask: how is a black hole different from a hot piece of coal? This central dogma is saying that, as long as you remain outside, it is not fundamentally different, in the sense that both are governed by a unitary Hamiltonian and have a finite number of degrees of freedom. Unlike a piece of coal, the black hole has an interior shrouded by an event horizon, and making it fully compatible with the exterior view is a non-trivial problem that has not been completely solved. 
\item
The name ``central dogma'' was borrowed from biology where the central dogma talks about the information transfer from DNA and RNA to proteins. Here it is also a statement about information  $-$ quantum information. It involves a certain dose of belief, because it is not something we can derive directly from the gravity description.  We can view it as an unproven assumption about the properties of a full theory of quantum gravity. It is also something that is not accepted by some researchers. In fact,  Hawking famously objected to it. 
\item
Notice that this statement is in stark contrast with a naive reading of the spacetime geometry. The spacetime geometry can be viewed as having two ``asymptotic'' regions. One is the obvious region outside, and the other is the future region near the singularity. See figure \ref{BabyUniverse}. Of course, the semiclassical gravity theory does not tell us how to evolve past this singularity, or even whether such evolution makes sense. From this point of view the interior is something we cannot access from the outside, but there is no obvious reason why some quantum information could not be lost here. In other words, if a black hole is a ``hole in space'' where things can get in and get lost, then the central dogma would be FALSE. In fact, this is one reason why some people think it is indeed false.  
\item
Both the results on black hole thermodynamics, as well as the results on fine-grained entropies we will discuss later, are true properties of a theory of gravity coupled to quantum fields and do {\it not} require the validity of the central dogma.   In other words, we are not assuming the central dogma in this review $-$ we are providing evidence for it. 
\end{itemize}

\subsection{Evidence from string theory for the central dogma}

Though we said that the ``central dogma'' is an unproven assumption, there is a great deal of very non-trivial evidence from string theory. 
String theory is a modification of Einstein gravity that leads to a well defined perturbative expansion and also some non-perturbative results. For this reason it is believed to define a full theory of quantum gravity. 

One big piece of evidence was the computation of black hole entropy for special extremal black holes in supersymmetric string theories \cite{Strominger:1996sh}. In these cases one can reproduce the 
 Bekenstein-Hawking   formula from an explicit count of microstates.  These computations match not only the area formula, but all its corrections, see e.g. \cite{Dabholkar:2014ema}. Another piece of evidence comes from the AdS/CFT correspondence \cite{Maldacena:1997re,Witten:1998qj,Gubser:1998bc}, which is a conjectured relation between the physics of AdS and a dual theory living at its boundary.  In this case, the black hole and its whole exterior can be represented in terms of degrees of freedom living at the boundary. There is also evidence from matrix models that compute scattering amplitudes  in special vacua \cite{Banks:1996vh}.  We will not discuss this further  in this review, since we are aiming to explain features which rely purely on gravity as an effective field theory. 

\section{Fine-grained vs coarse-grained entropy} \label{finecoarse}

There are two notions of entropy that we ordinarily use in physics and it is useful to make sure that we do not confuse them in this discussion. 

The simplest to define is the von Neuman entropy. Given the density matrix, $\rho$,  describing the quantum state of the system, we have 
\be \label{vnfine}
S_{vN} = - Tr[ \rho \log \rho ] 
\ee
 It quantifies our ignorance about the precise quantum state of the system.    It vanishes for a pure state, indicating complete knowledge of the quantum state. An important property is that it is invariant under unitary time evolution $\rho \to U \rho U^{-1}$.  

The second notion of entropy  is the coarse-grained entropy. Here we have some density matrix $\rho$ describing the system, but we do not measure all observables, we only measure a subset of simple, or coarse-grained observables $A_i$. Then the coarse-grained entropy is given by the following procedure. We consider all possible density matrices $\tilde \rho$ which give the same result as our system for the observables that we are tracking, 
$Tr[ \tilde \rho A_i] =Tr[\rho A_i]$. Then we compute the von Neumann entropy $S(\tilde \rho)$. Finally we maximize this over all possible choices of $\tilde \rho$. 

Though this definition looks complicated, a simple example is the ordinary entropy used in thermodynamics. In that case the $A_i$ are often chosen to be a few observables, say the approximate energy and the volume. The thermodynamic entropy is obtained by maximizing the von Neumann entropy among all states with that approximate energy and volume. 

Coarse-grained entropy obeys the second law of thermodynamics. Namely, it  tends to increase under unitary time evolution. 

Let us make some comments.
\begin{itemize}
	\item The von Neumann entropy is sometimes called the ``fine-grained entropy'', contrasting it with the  coarse-grained entropy defined above. Another common name is  ``quantum entropy." 
	\item 
 Note that the generalized entropy defined in \nref{sgen} increases rapidly when the black hole first forms and the horizon grows from zero area to a larger area.   Therefore if it has to be one of these two entropies, it can only be the thermodynamic entropy. In other words, the entropy \eqref{sgen} defined as the area of the horizon plus the entropy outside is the coarse-grained entropy of the black hole. 
\item 
Note that if we have a quantum system composed of two parts $A$ and $B$, the full Hilbert space is $H= H_A \times H_B$. Then we can define the von Neumann entropy for the subsystem $A$.  This is computed  by first forming a density matrix $\rho_A$ obtained by taking a partial trace over the system $B$. The entropy of $\rho_A$ can be non-zero, even if the full system is in a pure state. This arises when the original pure state contains some entanglement between the subsystems $A$ and $B$. 
In this case $S(A)=S(B)$ and $S(A\cup B) =0$. 
\item
The fine-grained entropy cannot be bigger than the coarse-grained entropy, $S_{vN} \leq S_{coarse}$. 
 This is a simple consequence of the definitions, since we can always consider $\rho$ as a candidate $\tilde \rho$. Another way to say this is that because $S_{coarse}$ provides a measure of the total number of   degrees of freedom available to  the system, it sets an upper bound on how much the system can be entangled with something else.

 \end{itemize}

\begin{figure}[h]
\begin{center}
\includegraphics[scale=.3]{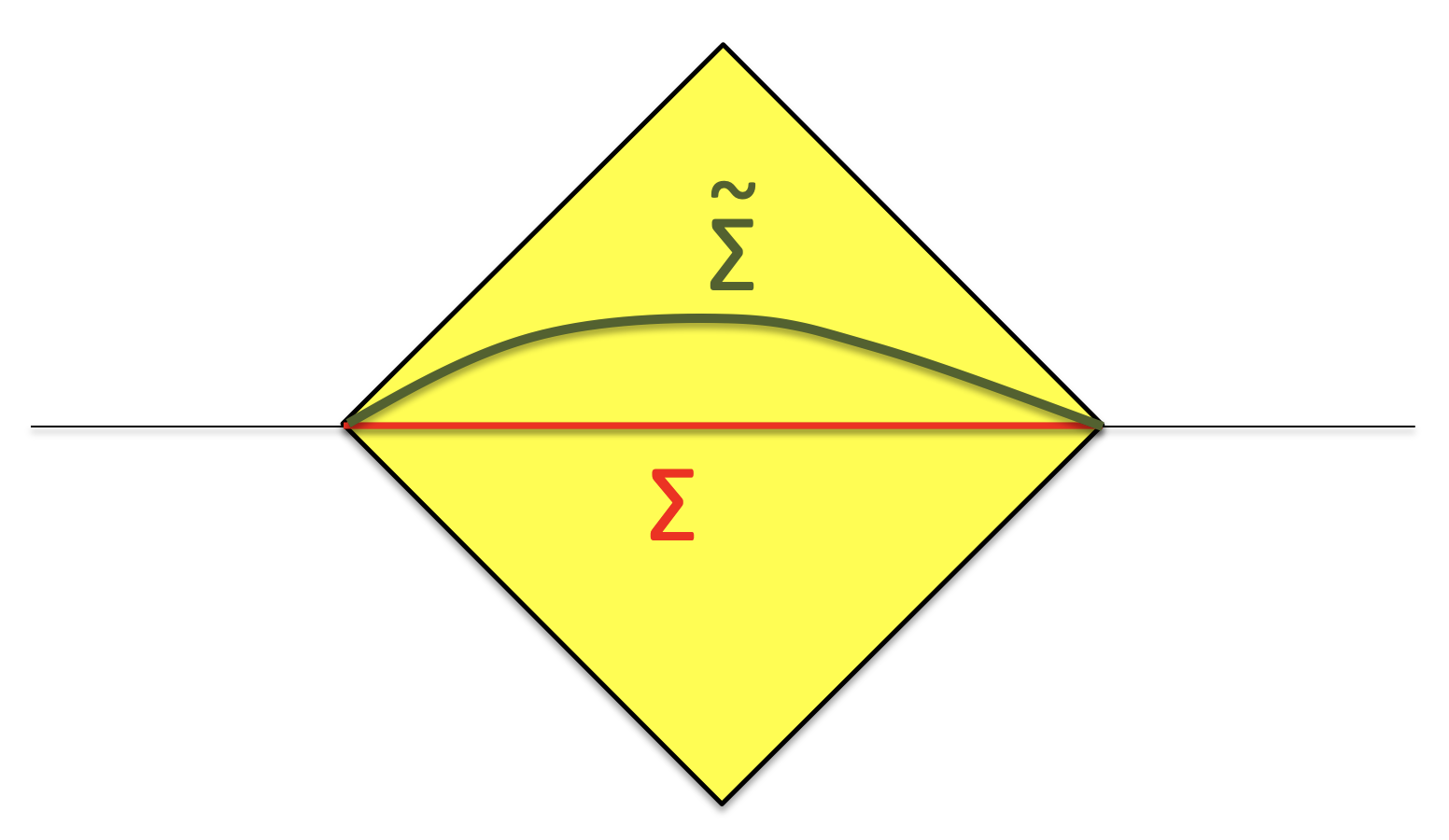}
\caption{ Given a region $\Sigma$ of a spatial slice, shown in red, we can define its causal diamond to be all points where the evolution is uniquely  determined by initial conditions on $\Sigma$. The alternative slice $\tilde \Sigma$ defines the same causal diamond. The von Neumann entropies are also the same.  }
\label{Diamond}
\end{center}
\end{figure}

It is useful to define the fine-grained entropy of the quantum fields in a region of space. Let $\Sigma$ be a spatial region, defined on some fixed time slice. This region has an associated density matrix $\rho_{\Sigma}$, and the fine-grained entropy of the region is denoted
\be
 S_{vN}(\Sigma) \equiv S_{vN}(\rho_\Sigma) \ .
\ee
If $\Sigma$ is not a full Cauchy slice, then we will have some divergences at its boundaries. These divergences are not important for our story, they have simple properties and we can deal with them appropriately.  Also, when $\Sigma$ is a portion of the full slice,   $S_{vN}(\Sigma)$ is generally time-dependent. It can increase or decrease with time as we move the slice  forwards   in time. The slice $\Sigma$ defines an associated causal diamond, which is the region that we can determine if we know initial data in $\Sigma$, but not outside $\Sigma$. The entropy is the same for any other slice $\tilde \Sigma$ which has the same causal diamond as $\Sigma$, see figure \ref{Diamond}.

\subsection{Semiclassical entropy}\label{semiclassical}

We now consider a gravity theory which we are treating in the semiclassical approximation. Namely, we have a classical geometry and quantum fields defined on that classical geometry. 
Associated to a spatial subregion we can define its 
 ``semiclassical entropy,"    denoted by
\be
\Ssemi(\Sigma) \ .
\ee
$\Ssemi$ is the von Neumann entropy of quantum fields (including gravitons) as they appear on the semiclassical geometry. In other words, this is the fine-grained entropy of the density matrix calculated by the standard methods of quantum field theory in curved spacetime. In the literature, this is often simply called the von Neumann entropy (it is also called $S_{\rm matter}$ or $S_{\rm outside}$ in the black hole context).

\section{The Hawking information paradox }

The   Hawking information paradox is an argument against the ``central dogma'' enunciated above \cite{Hawking:1976ra}. It is only a  paradox if we think that the central dogma is true.
 Otherwise, perhaps it can be viewed as a feature of quantum gravity.   

The basic point rests on an understanding of the origin of Hawking radiation. We can first start with the following question. Imagine that we make a black hole from the collapse of a pure state, such as a large amplitude gravity wave \cite{Christodoulou:2008nj}. This black hole emits thermal radiation. Why do we have these thermal aspects if we started with a pure state? The thermal aspects of Hawking radiation arise because we are essentially splitting the original vacuum state into two parts, the part that ends up in the black hole interior and the part that ends up in the exterior. The vacuum in quantum field theory is an entangled state. As a whole state it is pure, but the degrees of freedom are entangled at short distances. This implies that if we only consider half of the space, for example half of flat space, we will get a mixed state on that half. This is a very basic consequence of unitarity and relativistic invariance \cite{Bisognano:1976za}. 
 Often this is explained qualitatively as follows. The vacuum contains pairs of particles that are constantly being created and annihilated. In the presence of a horizon, one of the members of the pair can go to infinity and the other member is trapped in the black hole interior.  We will call them the ``outgoing Hawking quantum" and the ``interior Hawking quantum."  These two particles are entangled with each other, forming a pure state. However if we consider only one member, say the outgoing Hawking quantum, we fill find it in a mixed state, looking like a thermal state at the Hawking temperature \nref{thawking}. See figure \ref{fig:evap-stages}b and figure \ref{fig:evap-penrose}.

 This process on its own does not obviously violate the central dogma. In fact, if we had a very complex quantum system which starts in a pure state, it will appear to thermalize and will emit radiation that is very close to thermal. In particular, in the early stages, if we computed the von Neumann entropy of the emitted radiation it would  be almost exactly thermal because the radiation is entangled with the quantum system. So it is reasonable to expect that during the initial stages of the evaporation, the entropy of radiation rises. However, as the black hole evaporates more and more, its area will shrink and we run into trouble when the entropy of radiation is bigger than the thermodynamic entropy of the black hole. The reason is that now it is not possible for the entropy of radiation to be entangled with the quantum system describing the black hole because the number of degrees of freedom of the black hole is given by its thermodynamic entropy, the area of the horizon.
 In other words, if the black hole degrees of freedom together with the radiation  are producing a pure state, then the fine-grained entropy of the black hole should be equal to that of the radiation $S_{\rm black ~hole} = S_{\rm rad}$. But this fine-grained entropy of the black hole should be less than the Bekenstein-Hawking or thermodynamic entropy of the black hole, 
 $S_{\rm black~hole} \leq S_{\rm Bekenstein-Hawking}=S_{\rm coarse-grained}$. 
 
\begin{figure}[h]
\begin{center}
\includegraphics[scale=.5]{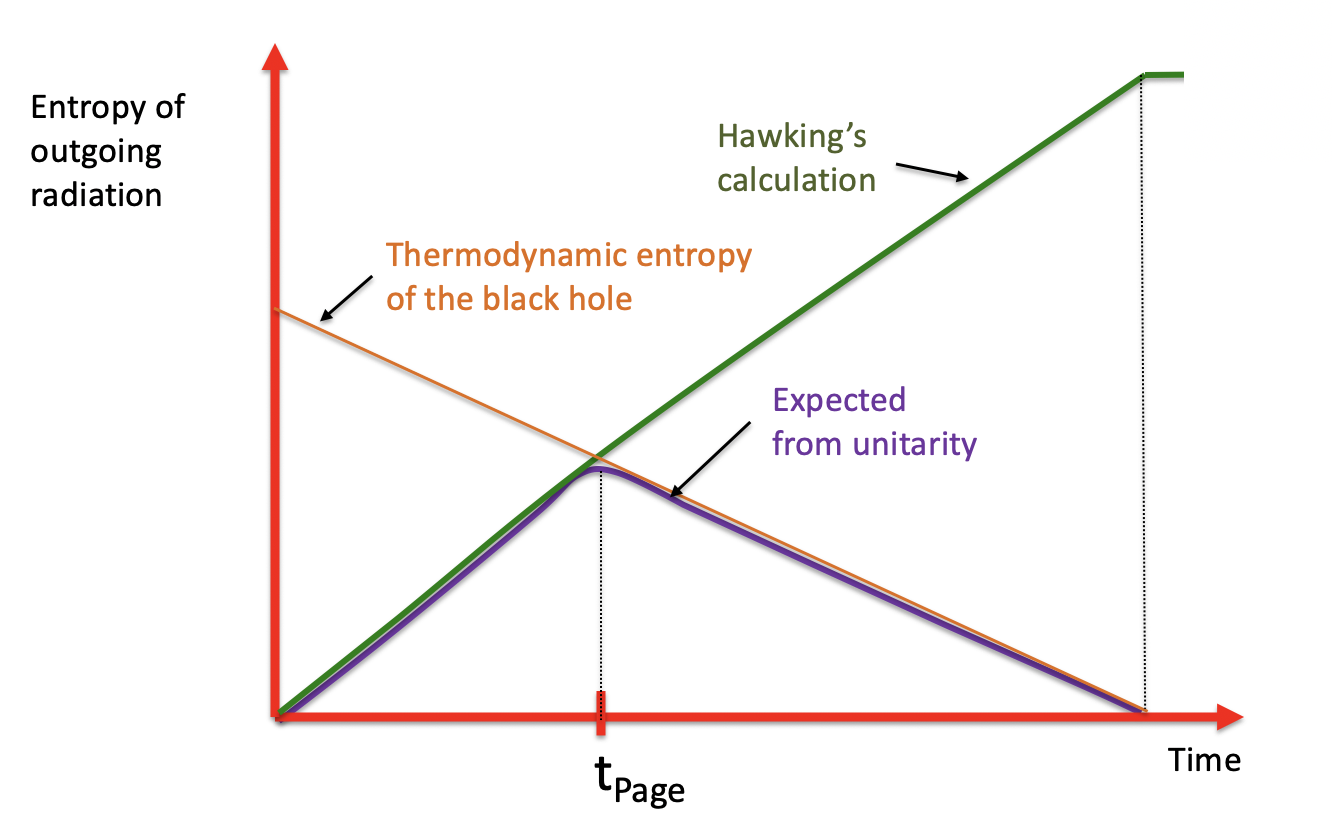}
\caption{  Schematic behavior of the entropy of the the outgoing radiation.  The precise shape of the lines depends on the black hole and the details of the matter fields being radiated.  In green we see Hawking's result, the entropy monotonically increases until $t_{\rm End}$, when the black hole completely evaporates. In orange we see the thermodynamic entropy of the black hole. If the process is unitary, we expect the entropy of radiation to be smaller than the thermodynamic entropy. If it saturates this maximum, then it should follow the so called ``Page'' curve, denoted in purple. This changes relative to the Hawking answer at the Page time, $t_{\rm Page}$, when the entropy of Hawking radiation is equal to the thermodynamic entropy of the black hole.  }
\label{HawkingPageCurves}
\end{center}
\end{figure}

 If the central hypothesis were true, we would expect that the entropy of radiation would need to start decreasing at this point. In particular, it can never be bigger than the Bekenstein Hawking entropy of the old black hole. Notice that we are talking about the von Neumann or fine-grained entropy of the radiation. 
 Then, 
 as suggested by D. Page \cite{Page:1993wv,Page:2013dx}, the entropy of the radiation would need to follow the curve indicated in figure \ref{HawkingPageCurves}, as opposed to the Hawking curve.
 The time at which $S_{\rm Bekestein-Hawking} = S_{\rm rad}$ is called the Page time.
 
 Now let us finish this discussion with a few comments. 
 
 \begin{itemize}
 \item
 Note that, as the black hole evaporates, its mass decreases. This is sometimes called the ``backreaction'' of Hawking radiation. This effect is included in the discussion. And it does not ``solve'' the problem.
 \item
 When the black hole reaches the final stages of evaporation, its size becomes comparable to the Planck length and we can no longer trust the semiclassical gravity description.  This is not relevant since the conflict with the central dogma appeared at the Page time, when the black hole was still very big. 
 \item The argument is very robust since it relies only on basic properties of the fine-grained entropy. In particular, it is impossible to fix the problem by adding small corrections to the Hawking process by slightly modifying the Hamiltonian or state of the quantum fields near the horizon \cite{Mathur:2009hf,Almheiri:2012rt,Almheiri:2013hfa}.  In other words, the paradox holds to all orders in perturbation theory, and so if there is a resolution it should be non-perturbative in the gravitational coupling $\GN$.
 \item
 We could formulate the paradox by constantly feeding the black hole with a pure quantum state	so that we exactly compensate the energy lost by Hawking radiation. Then the mass of the black hole is constant. Then the paradox would arise when this process goes on for a sufficiently long time that the entropy of radiation becomes larger than the entropy of the black hole. 
 \item
 One could say that the gravity computation only gives us an approximate description and we should not expect that a sensitive quantity like the von Neumann entropy should be exactly given by the  semiclassical theory. In fact, this is what was said until recently. We will see however, that there \emph{is} a way to compute the von Neuman entropy using just this semiclassical description. 
 \end{itemize}

 We have described here one aspect of the Hawking information paradox, which is the aspect that we will see how to resolve. We will comment about other aspects in the discussion.

\section{A formula for fine-grained entropy in gravitational systems}\label{finegrain}


As we mentioned above,  the Bekenstein Hawking entropy formula should be viewed as the coarse-grained entropy formula for the black hole, since it increases under time evolution. This is clear when the black hole first forms and has not yet had time to emit Hawking radiation. 

Surprisingly there is also a gravitational formula for the von Neumann or fine-grained entropy  \cite{Ryu:2006bv,Hubeny:2007xt,Faulkner:2013ana,Engelhardt:2014gca}. It is also given by a formula that involves a generalized entropy, with an area plus the entropy of fields outside. The only difference is in the choice of the dividing surface. Roughly the idea is that we choose a surface such that the generalized entropy is minimized. This minimal value is the fine-grained entropy,   
\be \la{Appr}
 S \sim {\rm min} \left[ {{ \rm Area } \over 4 \GN} + S_{\rm outside} \right]  \ .
 \ee 
 
 \begin{figure}[t]
\begin{center}
\includegraphics[scale=0.4]{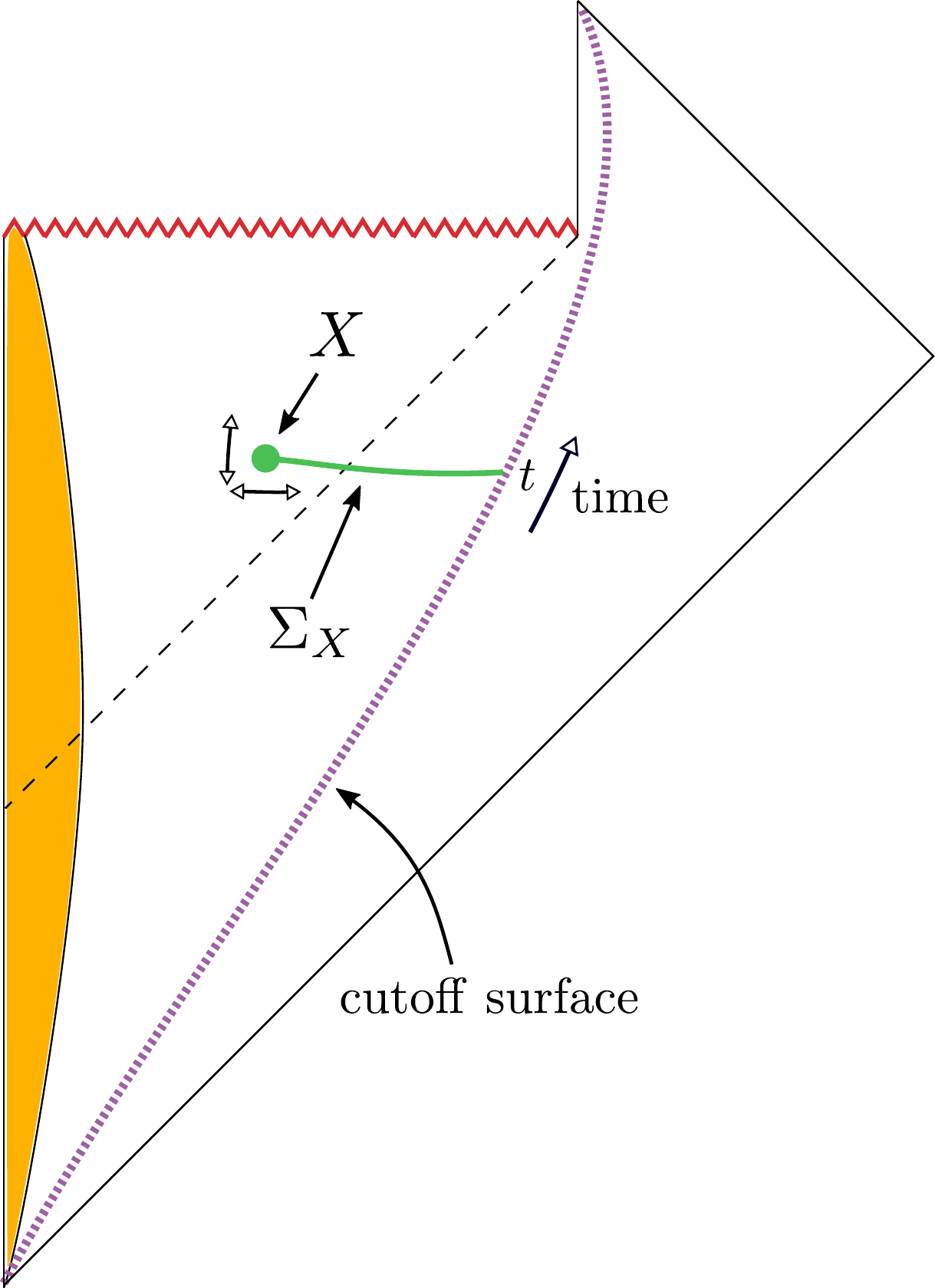}
\caption{The procedure for finding the extremal surface for computing the time-dependent fine-grained entropy of black hole. Starting on the cutoff surface at a given time, the surface $X$ is deformed into the enclosed region until an extremum is found.
Recall that this diagram represents the radial and time direction, so a point on this diagram represents a sphere.  }
\label{bhprocedure}
\end{center}
\end{figure}

 Now,  \nref{Appr} captures the spirit of the formula, but the precise formula is slightly more complicated. The reason is the following. A surface is a codimension-2 object. This means that it has two dimensions less than that of the full spacetime. In our case it is localized along one of the spatial dimensions and also in time. We are looking for a surface that     minimizes \nref{Appr} in the spatial direction but maximizes it in the time direction.  So we really should look for ``extremal surfaces'' by moving them both in space and in time. If there are many extremal surfaces we should find the global minimum. Another equivalent definition is the  following maxi-min construction  \cite{Wall:2012uf,Akers:2019lzs}.   First choose a spatial slice (a Cauchy slice) and find the minimal surface. Then find the maximum among all choices of the Cauchy slice.  Then a more precise version of the formula is 
 \cite{Ryu:2006bv,Hubeny:2007xt,Faulkner:2013ana,Engelhardt:2014gca} 
 \be \la{RT} 
\setlength\fboxsep{0.25cm}
\setlength\fboxrule{0.4pt}
 \boxed{ S= {\rm min}_X \left\{ {\rm ext}_X \left[ { {\rm Area}( X)  \over 4 G_N} + \Ssemi(\Sigma_X) \right] \right\}} \ ,
 \ee 
 where $X$ is a codimension-2 surface, $\Sigma_X$ is the region bounded by $X$ and the cutoff surface, and $\Ssemi(\Sigma_X)$ is the von Neumann entropy of the quantum fields on $\Sigma_X$ appearing in the semiclassical description, see figure \ref{bhprocedure}. The quantity in brackets is the generalized entropy,
 \be\la{sgendef}
 S_{\rm gen}(X) = { {\rm Area}( X)  \over 4 G_N} + \Ssemi(\Sigma_X) \ .
 \ee
 The idea is that we start from a surface outside the black hole and we can move it past the horizon into the interior to find the minimum. In particular this means that the answer depends on the geometry of the black hole interior. We can have black holes with similar exteriors but different interiors.  Such black holes will have different fine-grained entropies.
	It could happen that the surface can be shrunk completely in the interior of the black hole. In this case there is no area contribution. We will see this below in more detail. 
	
	\begin{figure}[t]
\begin{center}
\includegraphics[scale=0.4]{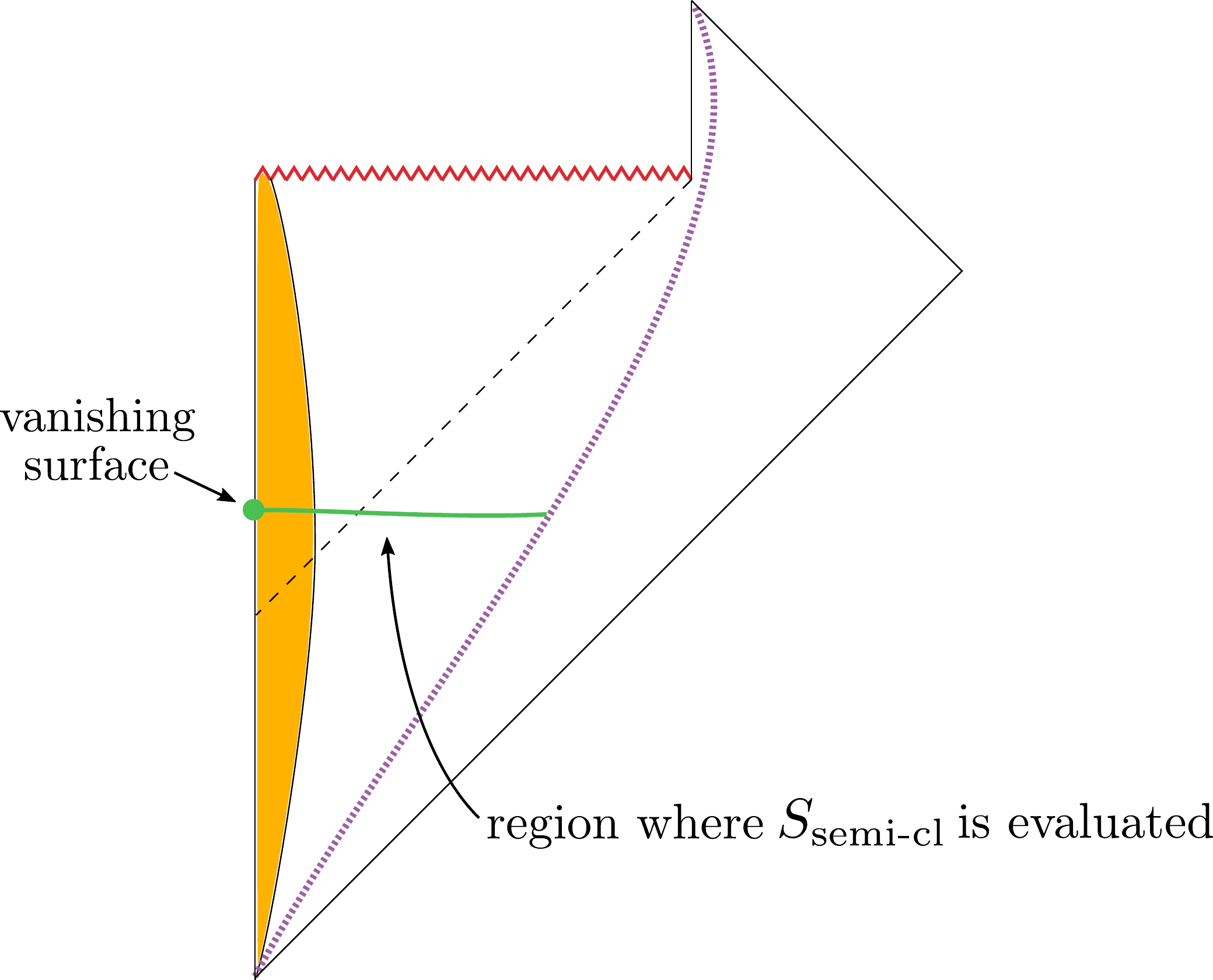}
\caption{  \small The minimal surface for the black hole at early times shrinks down to zero size. We call this the vanishing surface. The generalized entropy reduces to the bulk entropy of the entire region enclosed by the cutoff surface. \label{CollapseRT}}

\end{center}
\end{figure}

\begin{itemize}
		
	\item
In the literature the surface that extremizes \nref{RT} is called the ``quantum extremal surface." 
Note that this is just a classical geometric surface in the spacetime. It is called ``quantum'' because the matter  contribution in \nref{RT} contains the entropy of the quantum fields. 
\item
	 This formula, \nref{RT},  can be derived by a method similar to the Gibbons Hawking method discussed in section \ref{ss:gibbonshawking} when there is a path integral prescription for the construction of the state  \cite{Lewkowycz:2013nqa,Faulkner:2013ana,Dong:2017xht}.
	 Something similar will be discussed in section \ref{replicas}.
	
	\end{itemize}


\section{Entropy of an evaporating black hole}

In this section, we will see how to apply the fine-grained entropy  formula \eqref{RT} to all stages of the evaporating black hole.

	Let us first compute the entropy after the black hole forms but before any Hawking radiation has a chance to escape the black hole region.
	In this case, there are no extremal surfaces encountered by deforming $X$ inwards, and we are forced to shrink it all the way down to zero size.  See figure \ref{CollapseRT}. The area term vanishes, so the fine-grained entropy is just the entropy of the matter enclosed by the cutoff surface. 
	Note that this calculation is sensitive to the geometry in the interior of the black hole.  	This means that the entropy at the initial stage will vanish,  assuming that the collapsing shell was in a pure state.\footnote{   We are neglecting the contribution from the  entanglement of the fields near the cutoff surface. We are taking this contribution to be time independent, and we implicitly subtract it in our discussion.}   If we ignore the effects of Hawking radiation, this fine-grained entropy is invariant under time evolution. This is in contrast with the area of the horizon, which starts out being zero at $r=0$ and then grows to $4\pi r_s^2$ after the black hole forms.

\begin{figure}[h]
\begin{center}
\includegraphics[scale=0.39]{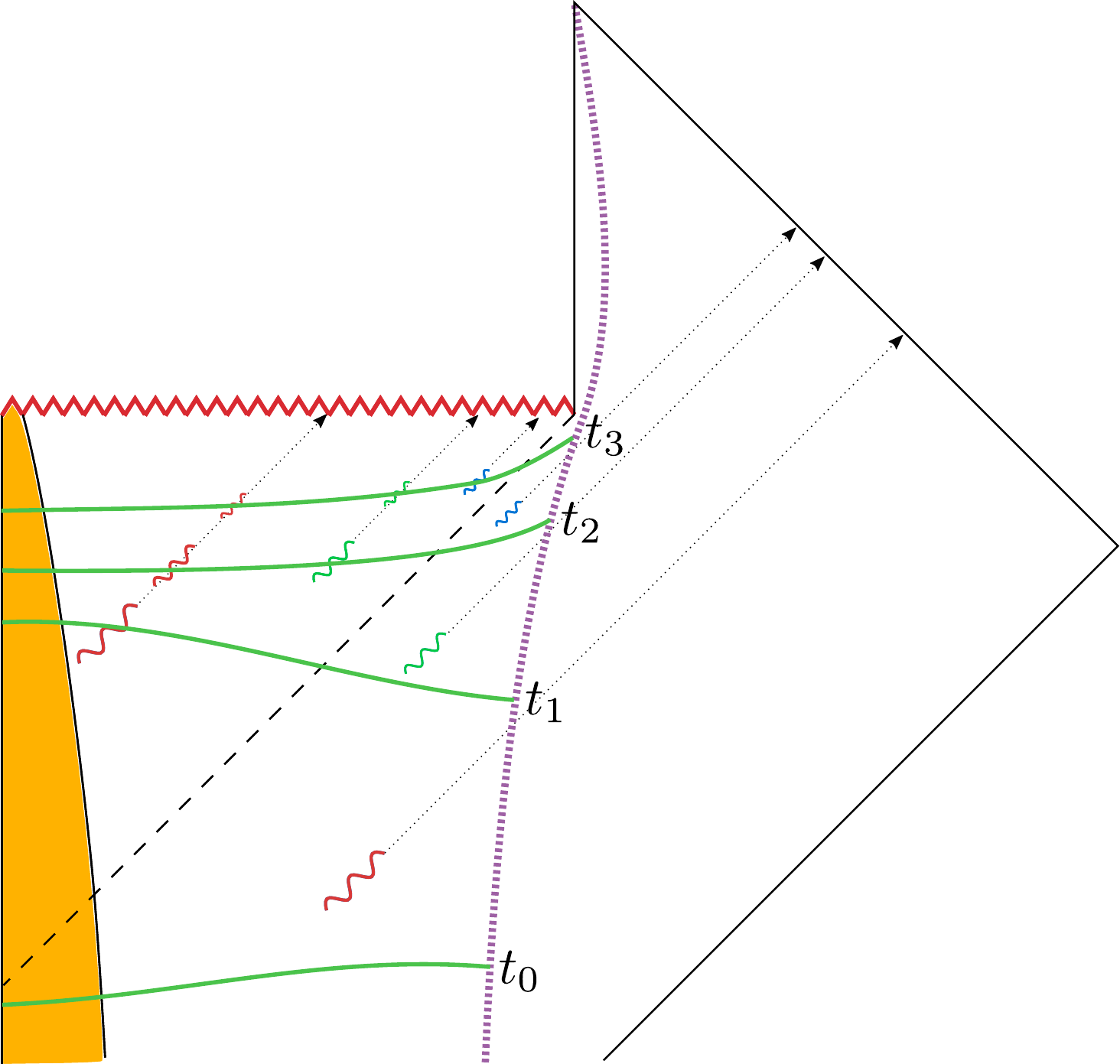}
 \ \ \ \ \ \ \includegraphics[scale=0.56]{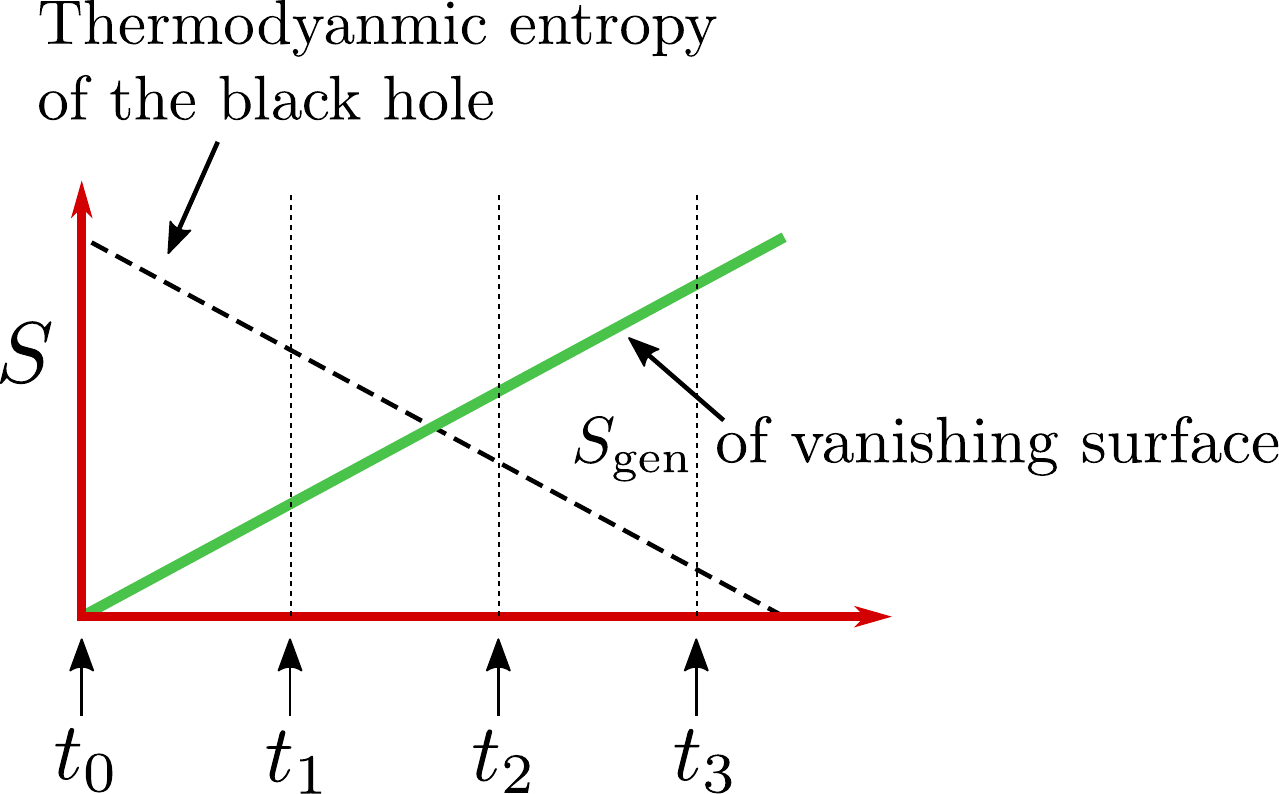}
\caption{As more outgoing Hawking quanta escape the black hole region, its entropy grows due to the pile up of interior Hawking quanta. Modes of like colors are entangled with one another. On the right is a plot comparing this growing entropy to the decreasing thermodynamic entropy of the black hole. }
\label{nicesliceentropy}
\end{center}
\end{figure}

Once the black hole starts evaporating and the outgoing Hawking quanta escape the black hole region, the von Neumann entropy of this region will no longer be zero due to the entanglement between the interior Hawking quanta and those that escaped. As shown in figure \ref{nicesliceentropy}, this entropy continues to grow as the black hole evaporates due to the pile up of the mixed interior modes. This growth of  entropy precisely parallels that of the outgoing Hawking radiation, and seems to support the idea that the black hole can have arbitrarily larger entropy than its Area$/4G_N$, inconsistent with the central dogma. 

\begin{figure}[t]
\begin{center}
\includegraphics[scale=0.39]{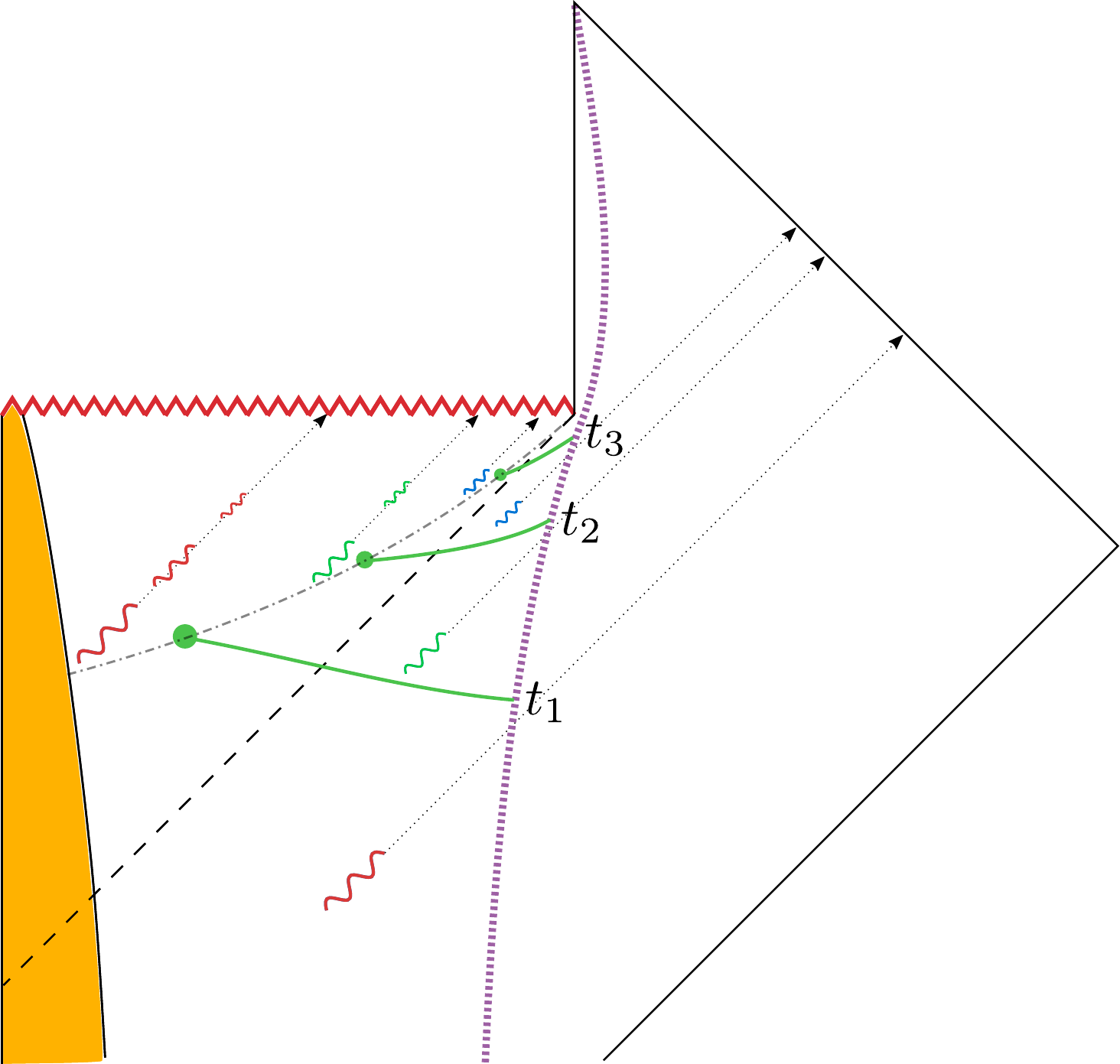}
 \ \ \ \ \ \ \includegraphics[scale=0.56]{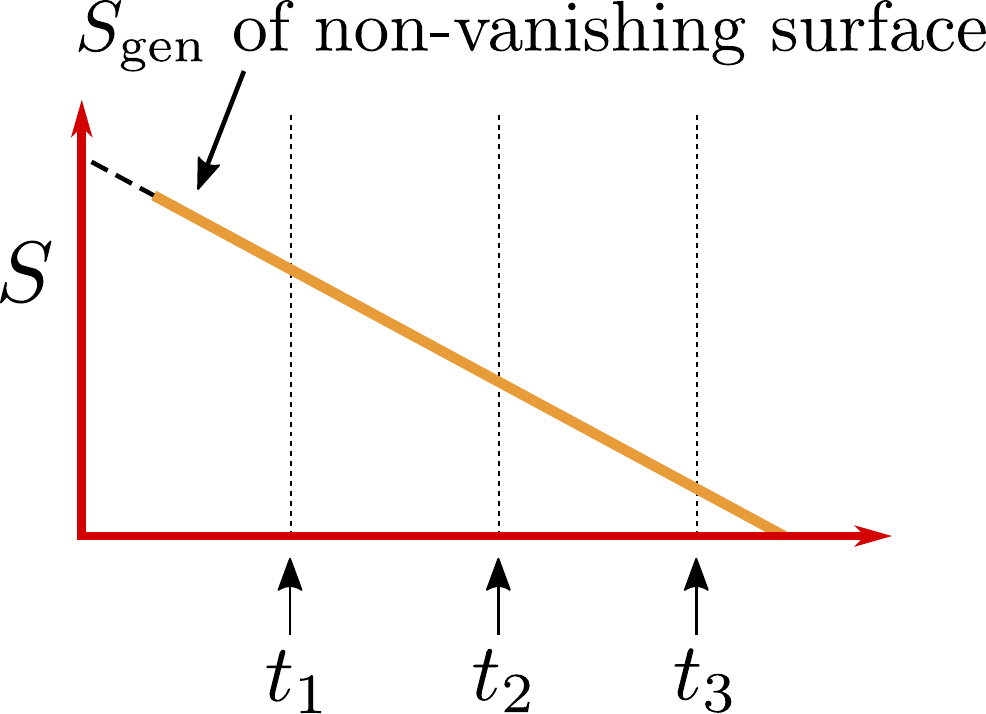}
\caption{When the non-vanishing extremal surface first appears, it lies inside the black hole near the event horizon. For different times on the cutoff surface, it is a different surface which moves along a spacelike direction up the horizon. This gives a decreasing generalized entropy since the black hole area is shrinking.}
\label{nontrivialsurface}
\end{center}
\end{figure}

The story is not yet complete since there is also a non-vanishing extremal surface that appears  shortly after the Hawking radiation starts escaping the black hole region. The precise location of this surface depends on how much radiation has escaped, and hence on the time $t$ along the cutoff surface when we decide to compute the entropy. 
It turns out that the surface lies close to the event horizon. 
Its location along the horizon is determined as follows. We go back along the cutoff surface by a time of order $r_s \log S_{BH}$ and we shoot an ingoing light ray. Then the surface is located close to the point where this light ray intersects the horizon. Note that $r_s$, and also $r_s \log S_{BH}$, are times which are short compared to the evaporation time, $r_s S_{BH}$. The time scale $r_s \log S_{BH}$ is called ``the scrambling time'' and it   has an interesting significance that we will not discuss in this review, see e.g. \cite{Hayden:2007cs,Sekino:2008he}. 
This is shown in figure \ref{nontrivialsurface}. The generalized entropy  now has an area term as well as the von Neumann entropy of quantum fields, $\Ssemi$. 
This quantum field theory entropy is relatively small because it does not capture many Hawking quanta and thus the  entropy is dominated by the area term
\begin{align}
S_\mathrm{gen} \approx {\mathrm{Horizon \ Area}(t)  \over 4 G_N} \,.
\end{align}
This generalized entropy follows closely the evolution of the thermodynamic entropy of the black hole. Since the area of the black hole decreases as it evaporates, this extremal surface gives a decreasing generalized entropy.

The complete proof for the existence of this surface would be to show that the change of area of $X$ under a small deformation in any direction perfectly balances the change in the von Neumann entropy  $\Ssemi$. We give some intuition for this procedure by extremizing along the ingoing null direction. The key point is that, while the area of $X$ is monotonically decreasing along this direction,  the entropy  $\Ssemi$ is not. To see this, imagine starting with $X$ right on the horizon and analyze the entanglement pattern across the surface as it is moved inwards. As the surface is moved inwards, the entropy $\Ssemi$   decreases as the newly included interior Hawking modes  purify the outgoing quanta already included in the region `outside.' Once all of those outgoing quanta have been purified, moving the surface inward further would start including modes entangled with outgoing quanta outside the black hole region thereby increasing $\Ssemi$. It is in this regime that changes in the area and entropy can exactly balance each other out. For the precise equations see \cite{Almheiri:2019psf,Penington:2019npb}.

\begin{figure}[t]
\begin{center}
\includegraphics[scale=.5]{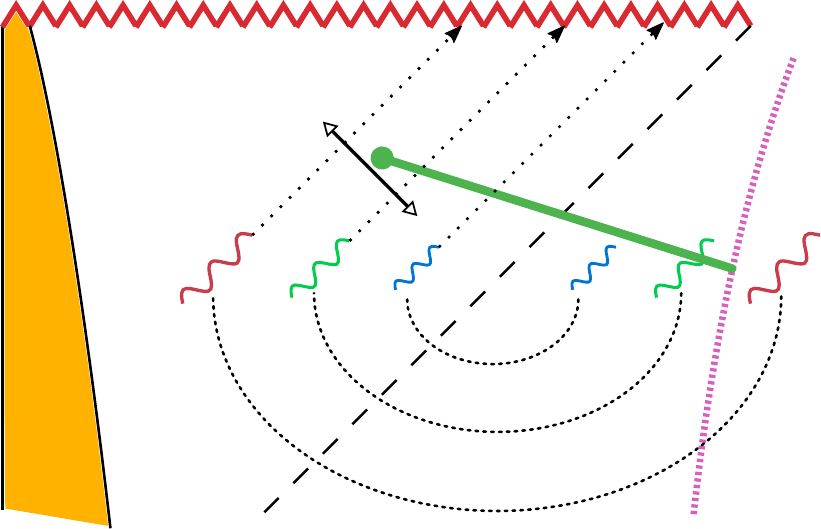}
\caption{$S_\mathrm{semi\text{-} cl}$ begins to increase when going inwards along the ingoing null coordinate once all the outgoing Hawking quanta in the black hole region are purified by the newly included interior modes. This allows for an extremum along this direction since the area of the surface shrinks.   }
\label{simpleextremization}
\end{center}
\end{figure}

\begin{figure}[t]
\begin{center}
\includegraphics[scale=0.39]{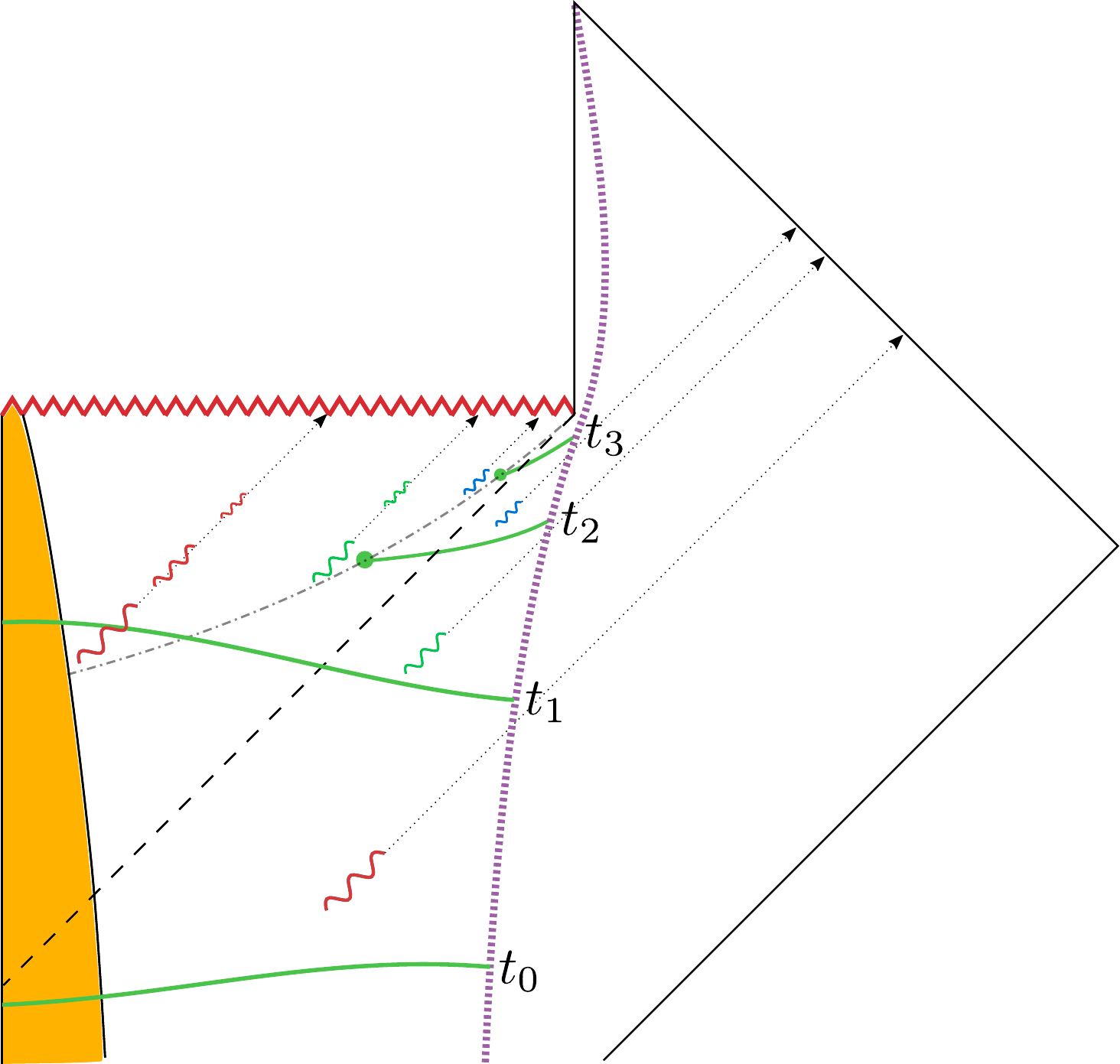}
 \ \ \ \ \ \ \includegraphics[scale=0.56]{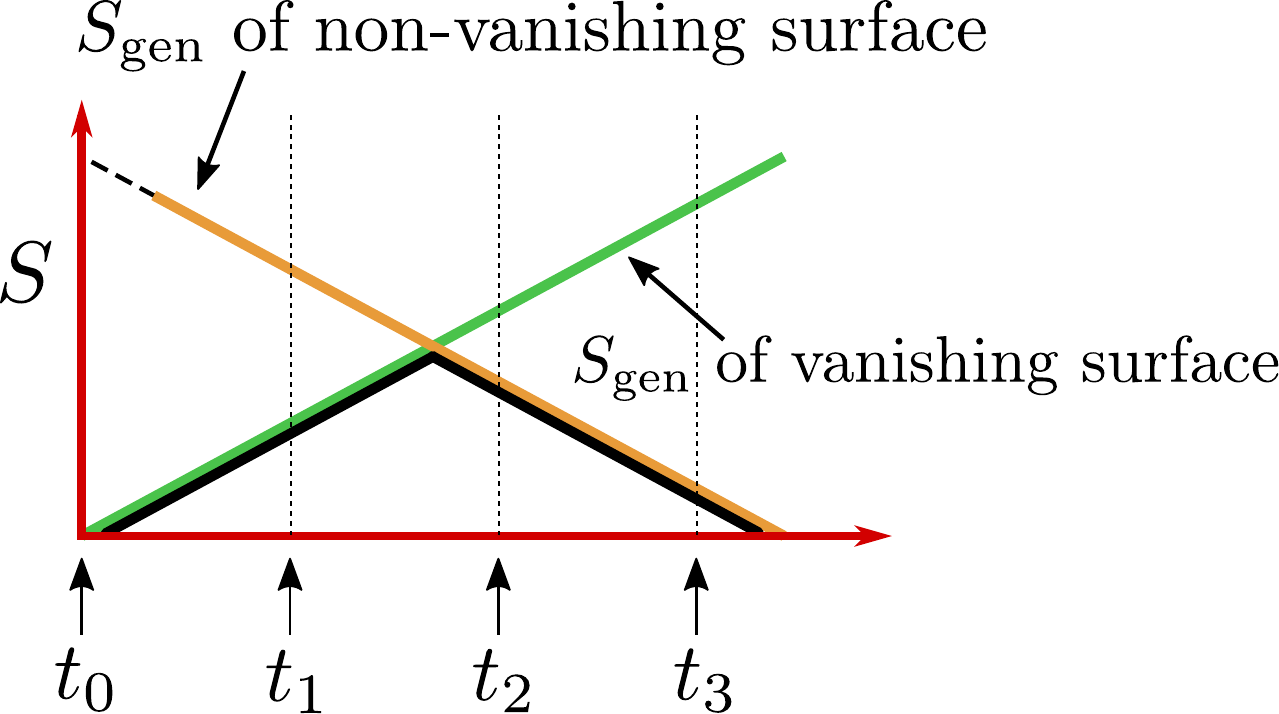}
\caption{The Page curve for the fine-grained entropy  of the black hole  (shown in black) is captured by the transition between a growing contribution from the trivial surface and a decreasing contribution from a non-trivial surface near the black hole horizon.}
\label{both}
\end{center}
\end{figure}

Full application of the entropy formula \eqref{RT} requires taking the minimum of the generalized entropy over all available extremal surfaces. We found two such surfaces: a growing contribution from the vanishing    surface and a decreasing one from the non-vanishing  surface just behind the event horizon. At very early times, only the vanishing surface exists, giving a contribution which starts at zero and grows monotonically until the black hole evaporates away. Some short time after the black hole forms, the non-vanishing surface is created and it starts with a large value given by the current area of the black hole, and decreases as the black hole shrinks. Therefore, the vanishing surface initially captures the true fine-grained entropy of the black hole, up until  the non-vanishing surface contribution becomes smaller and starts to represent the true fine-grained entropy. In this way, by transitioning between these two contributions, the entropy of the black hole closely follows the Page curve indicative of unitary black hole evaporation, see figure \ref{both}.

\section{Entropy of radiation}

\begin{figure}[t]
\begin{center}
\includegraphics[scale=0.4]{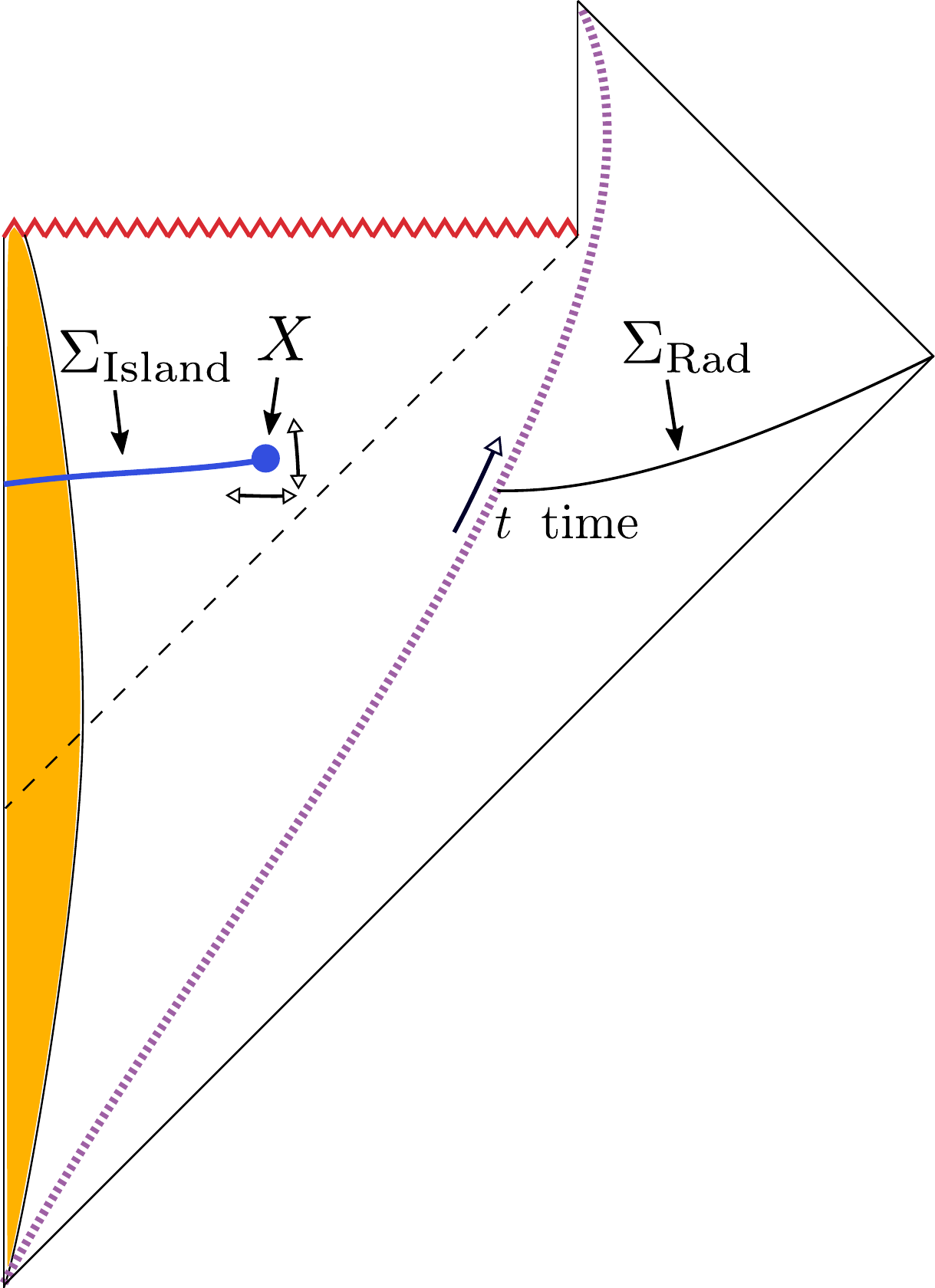}
\caption{The  fine-grained entropy of the region Rad containing the Hawking radiation can get contributions from regions inside the black hole, called islands. The total entropy is the area of $X$ plus a   contribution from the semiclassical entropy of the union of $\Sigma_{\rm Rad}$ and $\Sigma_{\rm Island}$.}
\label{islandprocedure}
\end{center}
\end{figure}

We have seen how the black hole   fine-grained entropy,  as computed via the gravitational formula \eqref{RT},  accurately follows the Page curve. This does not directly address the information paradox since that concerns the entropy growth of the Hawking radiation. 
In fact, the semi-classical black hole evolution leads to a growing value for the entropy outside the cutoff surface, the region containing the radiation \cite{Hawking:1976ra}, 
\be 
 \Ssemi(\Sigma_{\rm Rad})\,. \la{SemiRad}
 \ee 
This radiation lives in a spacetime region where the gravitational effects can be made very small. In other words, we can approximate this as a rigid space. Alternatively, we can think we collected the radiation into a big quantum computer.  
   However,   due to the fact that we used   gravity to obtain this state, it turns out that we should apply the gravitational  fine-grained entropy formula to compute its entropy.  
   In our first presentation of the formula \nref{RT}, we were imagining that we had a black hole inside the region. Now we are trying to apply this formula to the region outside the cutoff surface, which contains no black hole. Nevertheless, the formula \nref{RT} can also be applied when there is no black hole present.  The spirit of the formula is that the region in which we compute the entropy can be changed in size, by moving the surface $X$, so as to minimize the entropy. So far we have considered cases where $\Sigma_X$ was connected. However, it seems also natural to consider the possibility that $\Sigma_X$ is disconnected. When would this be required? By making $\Sigma_X$ disconnected, we increase the area of the boundary. So this can only be required if we can decrease the semiclassical entropy contribution at the same time. This could happen if we have regions that are far away with entangled matter. 
   In fact, this is precisely what happens with Hawking radiation. The radiation is entangled with the fields living in the black hole interior. Therefore, we can decrease the semiclassical entropy contribution by also including the black hole interior. In doing so, we will have to add an area term.   At late times, the net effect is to decrease the generalized entropy, so we are required to include this disconnected region inside the black hole, which is sometimes called an ``island." The final region we are considering looks as in figure \ref{islandprocedure}. 
   
   More precisely,  the full  fine-grained entropy of the radiation, computed using the  fine-grained gravitational entropy formula, is given by 
\begin{align}
S_\mathrm{Rad} = \mathrm{min}_X \Bigg\{ \mathrm{ext}_X\left[ {\mathrm{Area}(X) \over 4 \GN} + \Ssemi [\Sigma_{\rm Rad} \,  \cup \,  \Sigma_{\rm Island}] \right] \Bigg\}, \label{island}
\end{align}
where the area here refers to the area of the boundary of the island, and the min/ext is with respect to the location and shape of the island \cite{Almheiri:2019hni,Penington:2019kki,Almheiri:2019qdq} . 
The left hand side is the full entropy of radiation.  
 And $\Ssemi [\Sigma_{\rm Rad} \,  \cup \,  \Sigma_{\rm Island}]    $ is the von Neumann entropy of the quantum state of the combined radiation and island systems {\it in the semiclassical description}. 
Note that the subscript `Rad' appears both on the left hand side and the right hand side of \nref{island}, a fact that has caused much confusion and heated complaints. The left hand side is the full entropy of radiation, as computed using the gravitational  fine-grained entropy formula. This is supposed to be the entropy for the full exact quantum state of the radiation.  On the right hand side we have the state of radiation {\it in the semiclassical description}. This is a different state than the full exact state of the radiation. Note that in order to apply the formula we do not need to know the exact state of the radiation. The formula does not claim to give that state  to us in an explicit form, it is only computing the entropy of that state.

The ``island formula'' \nref{island} is a generalization of the black hole gravitational  fine-grained entropy formula \nref{RT} and really follows from the same principles. Some authors consider it as simply a part of \nref{RT}. We decided to give it a special name and to discuss it separately because we motivated \nref{RT} as a generalization of black hole entropy. However, if we look just at the radiation, there does not seem to be any black hole.  The point is that because we prepared the state using gravity, this is the correct formula to use. We will later give a sketch of the derivation of the formula. It is likely that in future treatments of the subject, both will be discussed together.

The procedure for applying this formula is as follows. We want to compute the entropy of all of the Hawking radiation that has escaped the black hole region. This is captured by computing the entropy of the entire region from the cutoff all the way to infinity. This region is labeled by $\Sigma_{\rm Rad}$  in the formula, see figure \ref{islandprocedure}. The islands refer to any number of regions contained in the black hole side of the cutoff surface. The figure shows the case of a single island centered around the origin. In principle we can have any number of islands, including zero. We then extremize the right hand side of 
\nref{island} with respect to the position of the surface $X$. Finally we minimize over all possible extremal positions and choices of islands. 

\begin{figure}[t]
\begin{center}
\includegraphics[scale=0.4]{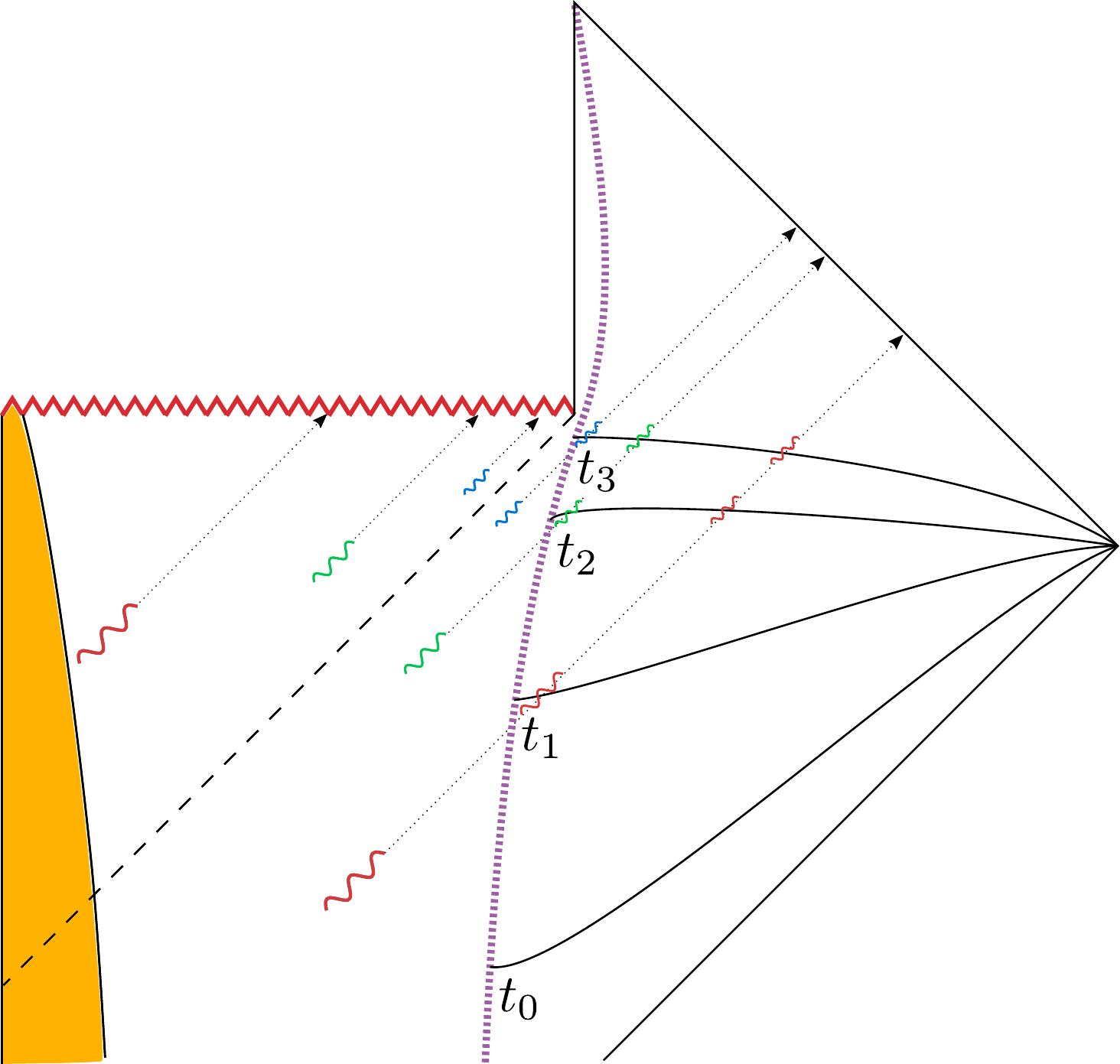}
 \ \ \ \ \ \  \includegraphics[scale=0.6]{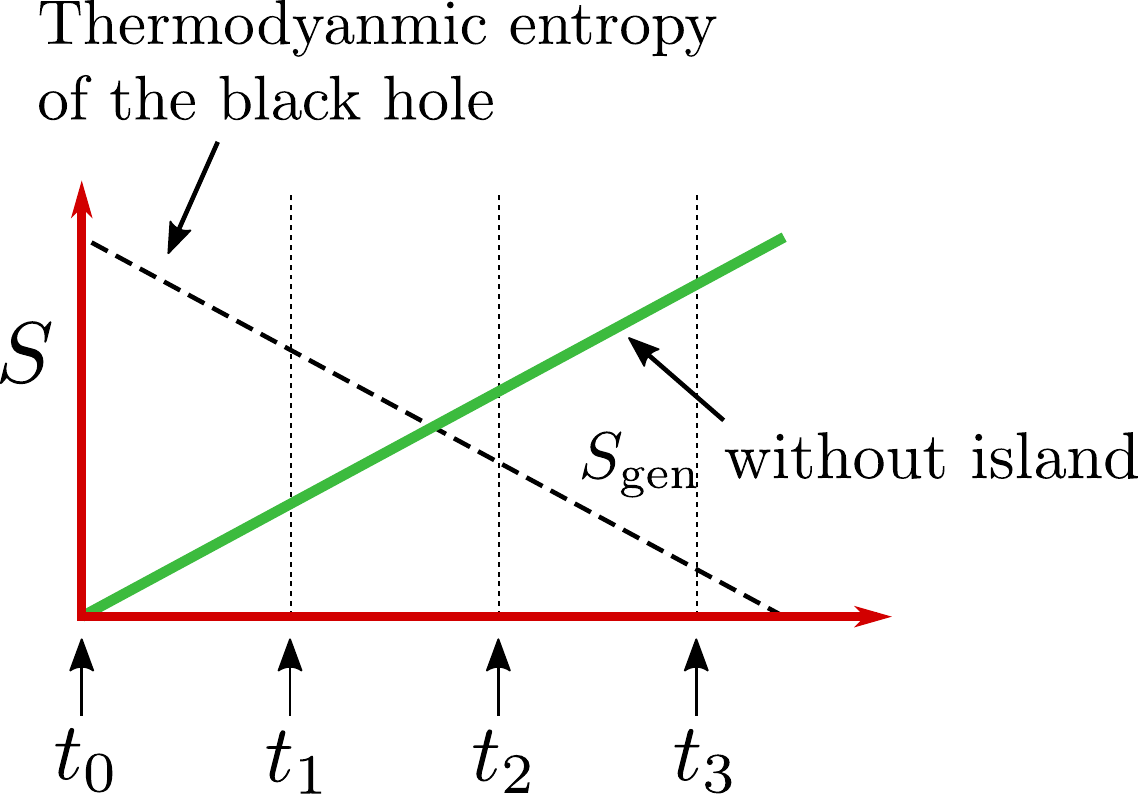}
\caption{The no-island contribution to the island formula gives a growing entropy due to the escaping outgoing Hawking quanta.  }
\label{radnoisland}
\end{center}
\end{figure}

The simplest possibility is to have no island. 
This vanishing island contribution gives simply \nref{SemiRad}.   As more and more outgoing Hawking quanta escape the black hole region, the entropy continues to grow. See figure \ref{radnoisland}. This contribution always extremizes the generalized entropy but it will not always be the global minimum of the entropy. 

\begin{figure}[htbp]
\begin{center}
\includegraphics[scale=0.4]{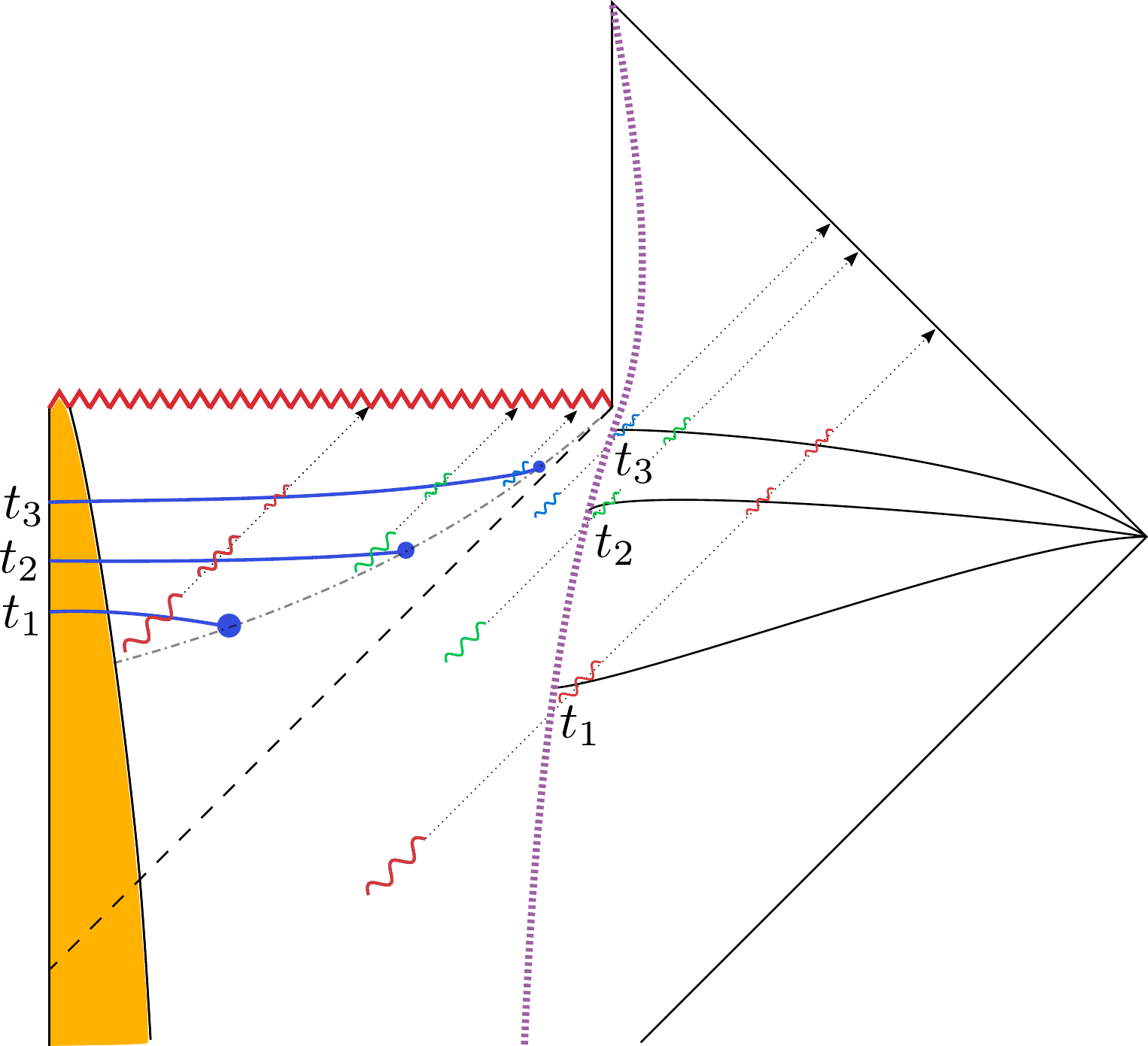}
 \ \ \ \ \ \  \includegraphics[scale=0.6]{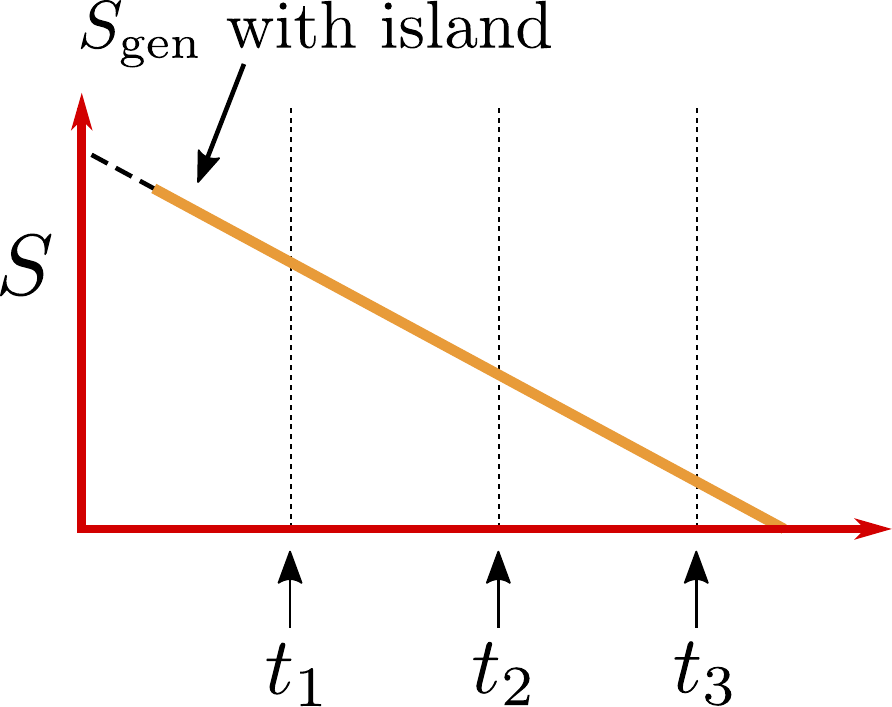}
\caption{The  island contribution appears some time after the black hole forms. It gives a decreasing contribution that tracks the thermodynamic entropy of the black hole.   }
\label{radwithisland}
\end{center}
\end{figure}

A non-vanishing island that extremizes the generalized entropy appears some time after the black hole forms. A time of order $r_s \log S_{BH}$ is enough. This island is centered around the origin and its boundary is very near the black hole event horizon. It moves up the horizon for different times on the cutoff surface. This is shown in figure \ref{radwithisland}. The generalized entropy with this island includes the term which is given by the area of the black hole. The von Neumann entropy term involves the entropy of the union of the outgoing radiation and the island, and is therefore small for all times, since the island contains most or all of the interior Hawking modes that purify the outgoing radiation. This contribution to the island formula starts at a large value, given by the area of the horizon at early times, and decreases down to zero.

\begin{figure}[htbp]
\begin{center}
\includegraphics[scale=0.4]{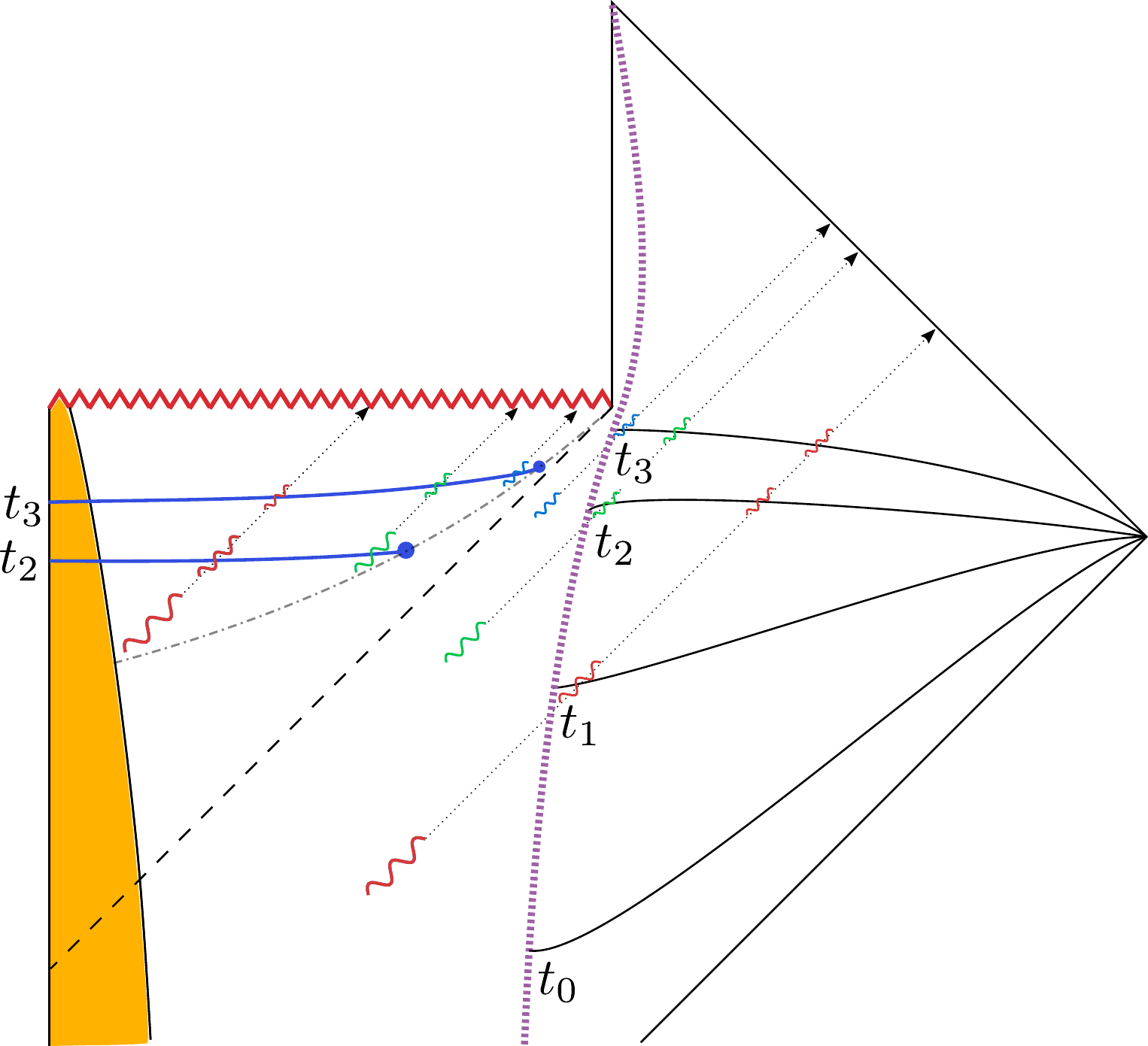}
 \ \ \ \ \ \ \includegraphics[scale=0.6]{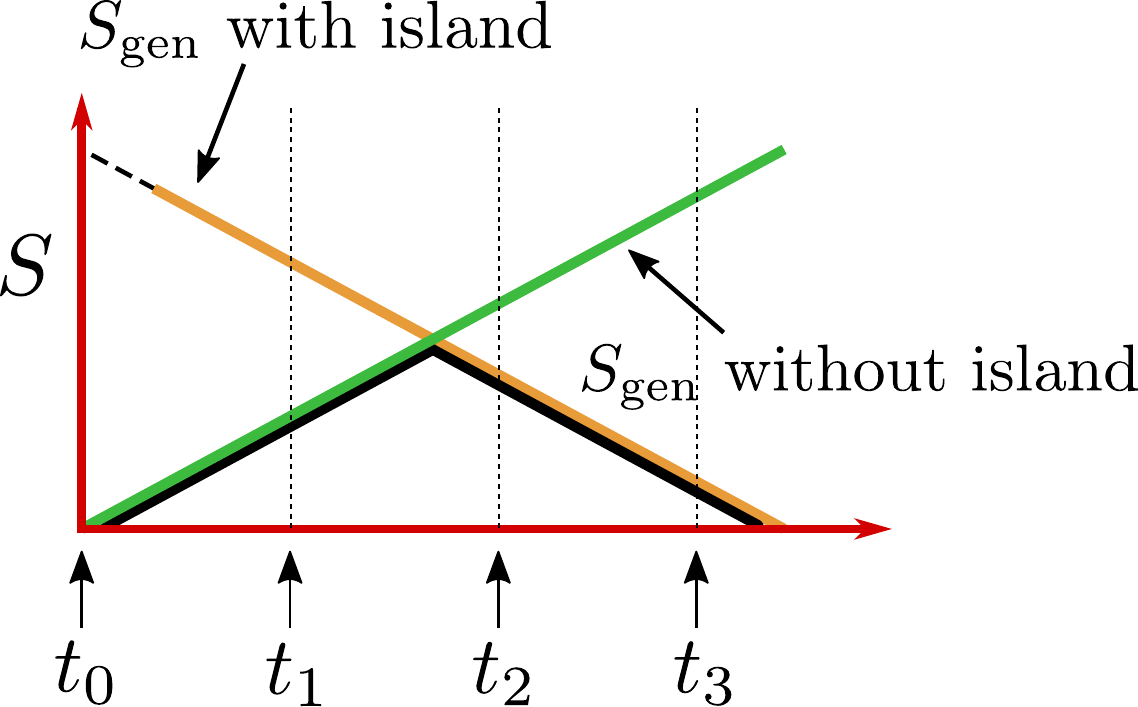}
\caption{ We now consider both contributions and at each time we pick the minimal one, which gives the final answer for the full entropy of radiation.   This gives the Page curve, shown in black on the right figure.    }
\label{radboth}
\end{center}
\end{figure}

The  fine-grained entropy of the Hawking radiation is then   the minimum of these two contributions. This gives the Page curve; the rising piece comes from  the no-island contribution and the novel decreasing piece from the island contribution.

If we formed the black hole from an initially pure state, then we expect that the entropy of the black hole and the entropy of the radiation  region should be equal. 
 Indeed, the fine grained entropy formula for the black hole and the one for the radiation give the same answer. The reason is the following. In both cases,  the same surface $X$ is involved. In addition, When the matter state is pure on the whole Cauchy slice,  we have that 
  $\Ssemi(\Sigma_X) = \Ssemi(\Sigma_\textup{Rad} \cup \Sigma_\textup{Island})$. Then we get the same answer because we are minimizing/extremizing the same function. 
 In conclusion, the black hole and the radiation entropy are described by the same curve, see figure \ref{HawkingPageCurves}.

Now a skeptic would say: ``Ah, all you have done is to include the interior region. As I have always been saying, if you include the interior region you get a pure state,"  or ``This is only an accounting trick." But we did not include the interior ``by hand." The  fine-grained entropy formula is derived from the gravitational path integral through a method conceptually similar to the derivation of the black hole entropy by Gibbons and Hawking, discussed in section \ref{ss:gibbonshawking}. It is gravity itself, therefore, that instructs us to include the interior in this calculation. 
It is gravity's way of telling us that black hole evaporation is unitary without giving us the details of the actual state of the outgoing radiation.

An analogy from the real world is the following. Imagine that there is a man who owns a house with many expensive paintings inside. Then he starts borrowing money from the townspeople giving the paintings as a guarantee. He spends his money throwing expensive parties and people who visit his house think he is very rich. However, he eventually borrows so much that most of the house and its contents belong to the townspeople. So, when the townspeople  compute their wealth, they include the paintings in this man's house. But the man cannot include them in his computation of his wealth. 
In this analogy, the house is the interior of the black hole and the wealth is the quantum information. The townspeople is the region far from the black hole containing the Hawking radiation.  The casual observer who thinks that the townspeople are poor because they don't have paintings in their homes would be wrong.  In the same way,  the one who looks at the Hawking radiation and said that it is mixed would be wrong, because the interior should also be included.


\section{The entanglement wedge and the black hole interior} \label{wedge}

 The central dogma talks about some degrees of freedom which suffice to describe the black hole from the outside. A natural question to ask is whether these degrees of freedom also describe the interior.   We have several possibilities: 
 
 a) They do not describe the interior. 
 
 b) They describe a portion of the interior. 
 
 c) They describe all of the interior. 
 
A guiding principle has been the formula for the fine-grained entropy of the black hole. This formula is supposed to give us the entropy of the density matrix that describes the black hole from the outside, if we are allowed to make arbitrarily complicated measurements. 
 We have seen that the answer for the entropy depends on the geometry of the interior. However, it only depends on the geometry and the state of the quantum fields up to the extremal   surface. Note that if we add an extra spin in an unknown state between the cutoff surface and the extremal  surface, then it will modify the fine-grained entropy. 
 
 Therefore it is natural to imagine that the degrees of freedom in the central dogma describe the geometry up to the minimal surface. If we know the state on any spatial region, we also know it in the causal diamond associated to that region, recall figure \ref{Diamond}. This has historically been called the ``entanglement wedge" \cite{Czech:2012bh,Wall:2012uf,Headrick:2014cta}. Following our presentation perhaps a better name would be ``the fine-grained entropy region," but we will not attempt to change the name. 

\begin{figure}[ht]
\begin{center}
\includegraphics[scale=.35]{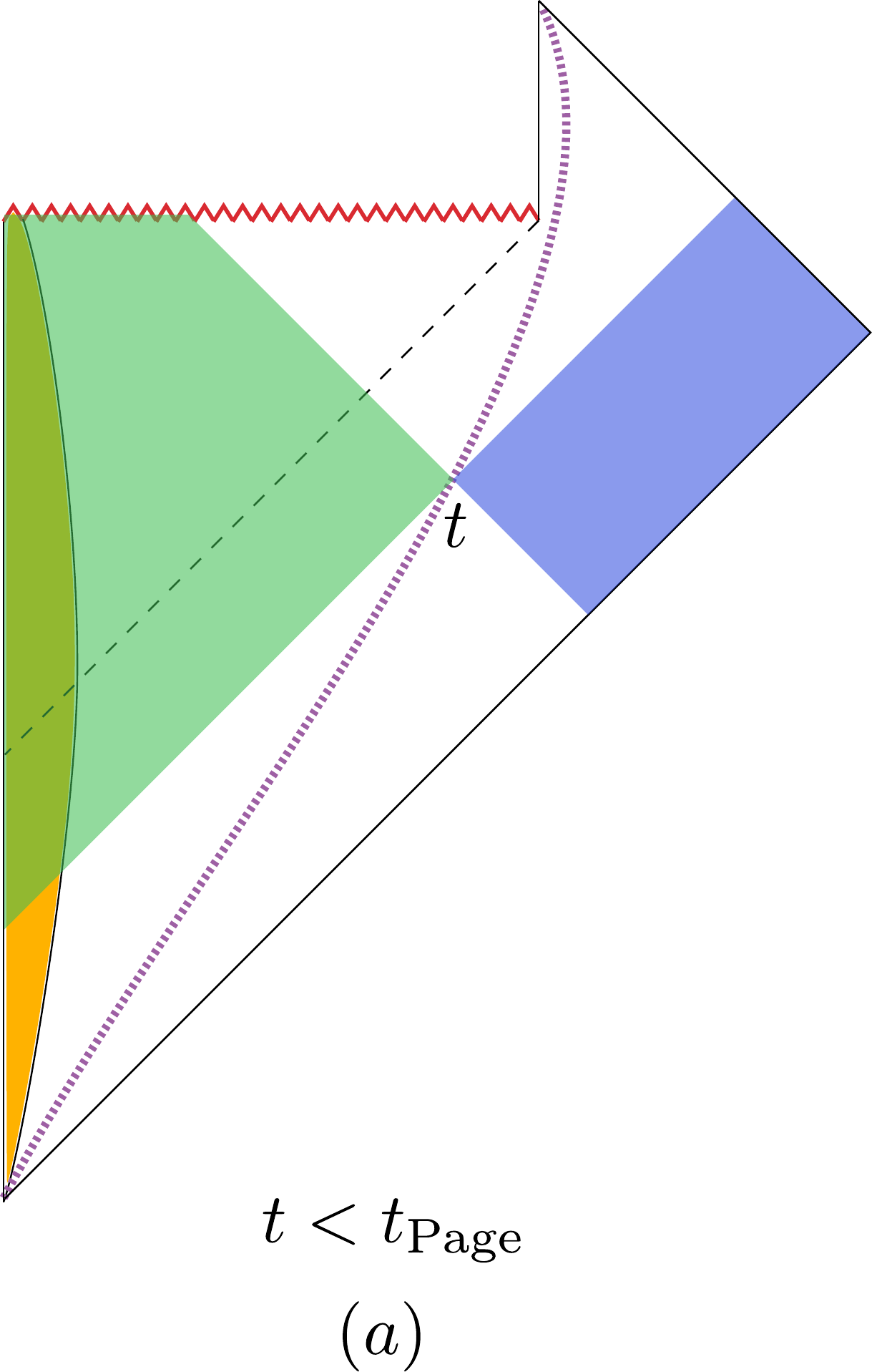} \ \ \ \ \ \  \ \
\includegraphics[scale=.35]{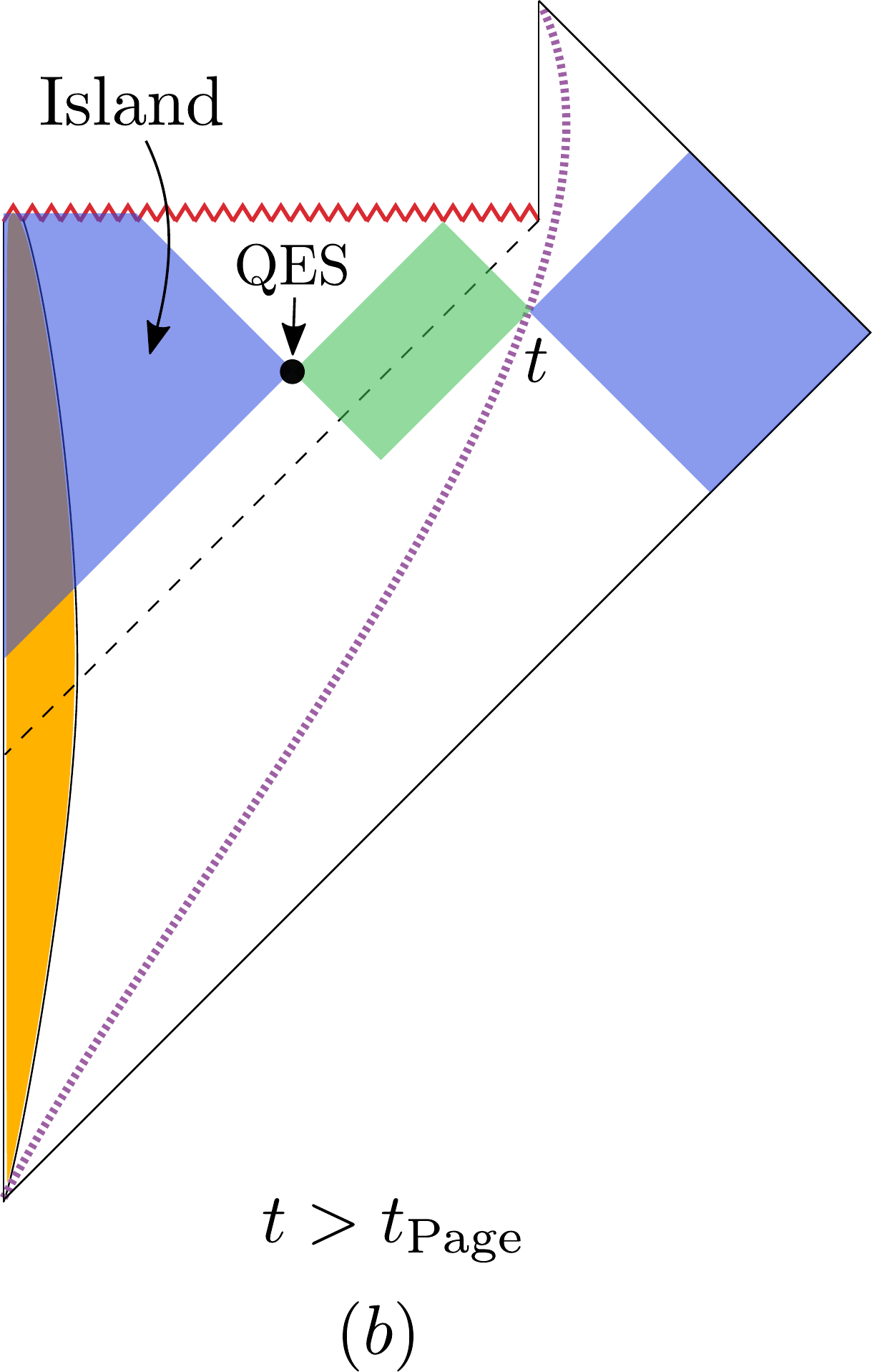}  \ \ \ \ \ \  \ \
\includegraphics[scale=.35]{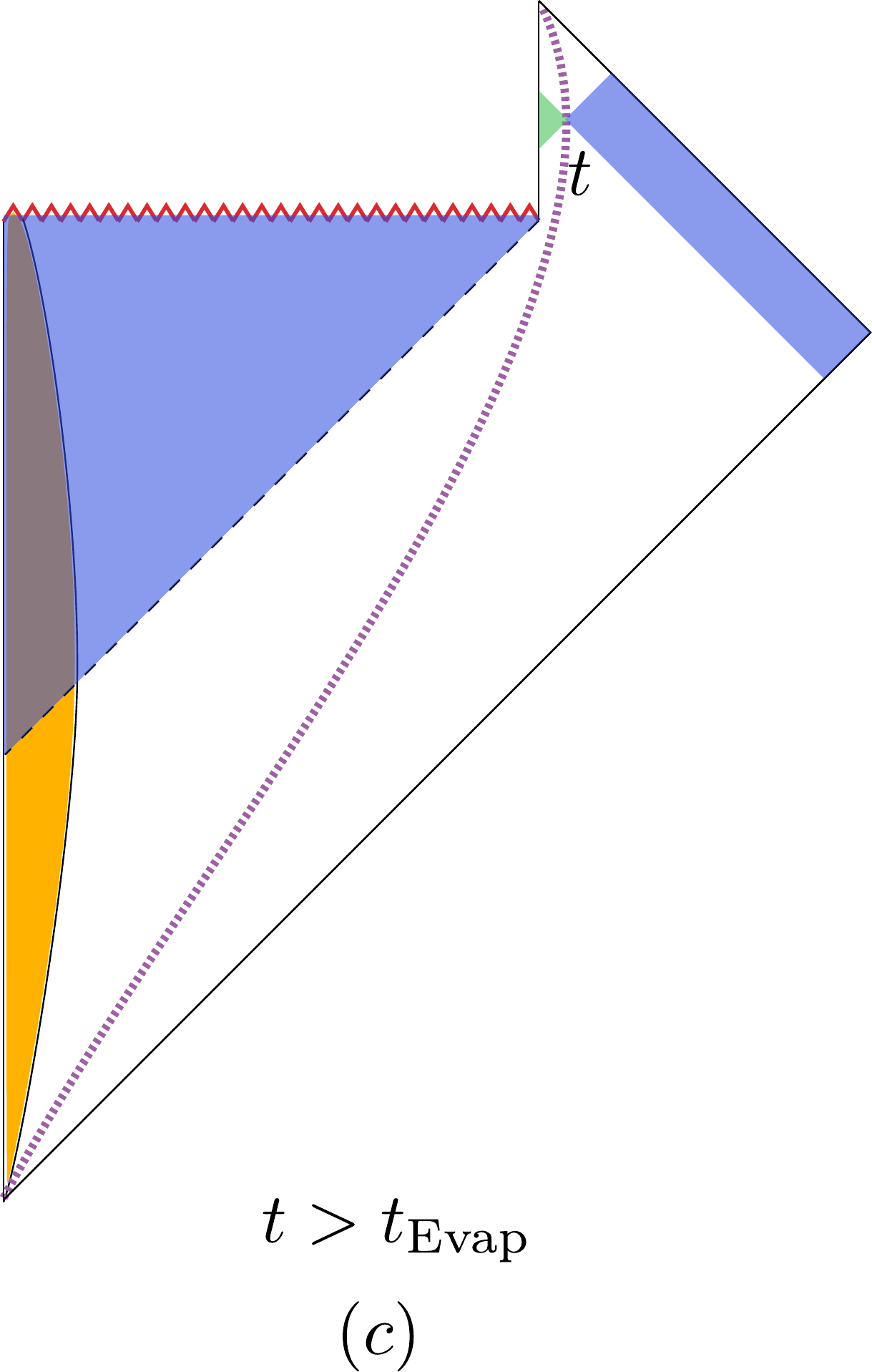} 
\caption{ In green we show the entanglement wedges of the black hole and in blue the entanglement wedges of the radiation region. Different figures show the wedges at different times. They are different because there is transfer of quantum information through the cutoff surface. To describe the white regions we need information both from the black hole region and the radiation region.   }
\label{EWfig}
\end{center}
\end{figure}

As a first example, let us look again at a black hole formed from collapse but before the Page time. The minimal surface is now a vanishing surface at the origin and the entanglement wedge of the black hole is the region depicted in green in figure  \ref{EWfig}a.   

As a second   example, we can look at the entanglement wedges of both the black hole and the radiation at late times, larger than the Page time. These are shown in figure \ref{EWfig}(b). 
The idea is that the black hole degrees of freedom describe the region of the spacetime in the black hole entanglement wedge while the radiation describes the degrees of freedom in the radiation entanglement wedge. It is important that the degrees of freedom that describe the black hole only describe a portion of the interior, the green region in figure \ref{EWfig}(b). The rest of the interior is encoded in the radiation.

Note how this conclusion, namely that the interior belongs to the entanglement wedge of the radiation, follows from the same guiding principle of using the fine-grained entropy. Since the fine-grained entropy of the radiation after the Page time contains the interior as part of the island, its entropy is sensitive to the quantum state of that region; a spin in a mixed state in the island contributes to the fine-grained entropy of the radiation.

Finally, as a third example,  we can consider a fully evaporated black hole, see figure \ref{EWfig}(c). In this case the region inside the cutoff surface is just flat space. The entanglement wedge of the radiation includes the whole black hole interior. This picture assumes that nothing too drastic happens at the endpoint of the evaporation.

 So far we have been a bit vague by the statement that we can ``describe'' what is in the entanglement wedge. A more precise statement is the ``entanglement wedge reconstruction hypothesis," which says that if we have a relatively small number of qubits  in an unknown state but located inside the entanglement wedge of the black hole, then by performing   operations on the black hole degrees of freedom, we can read off the state of those qubits. 
   This  hypothesis is supported by general principles of quantum information. Let us consider the case when the radiation entanglement wedge covers most of the interior, as in figure \ref{EWfig}(b).  Then the action of the  black hole  interior  operators of the semiclassical description   affect the entropy of radiation, according to the gravitational entropy formula. Assuming that this formula captures  the true entropy of the exact state of radiation, this means that these operators are changing this exact state \cite{Jafferis:2015del,Dong:2016eik}, see also \cite{Almheiri:2014lwa}.   Then it follows from general quantum information ideas that there is a map, called the 
   Petz map \cite{PetzBook},  that allows us to recover the information \cite{Cotler:2017erl}. In the context of simple gravity theories, this map can be constructed using the gravitational path integral \cite{Penington:2019kki}, again via the replica method. This provides a formal argument, purely from the gravitational side, for the validity of the hypothesis. 
    The actual quantum operations we would need to perform on the radiation are expected to be exceedingly complex, with a complexity that is (roughly) exponential in the black hole entropy 
    \cite{Brown:2019rox,Kim:2020cds}.
  
 For black holes after the Page time,  most of the interior is {\it not} described by the black hole degrees of freedom appearing in the central dogma. In fact, it is described by the radiation degrees of freedom. At late times, these are much more numerous than the area of the horizon.

	Note that there is an unfortunate language problem which sometimes gets in the way of the concepts we are trying to convey. The reason is that there are two {\it different} things that people might want to call ``black hole degrees of freedom." We have been calling ``black hole degrees of freedom'' the ones that appear in the central dogma. They are the ones that are sufficient to describe the black hole from the outside. These are not very manifest in the gravity description. The other possible meaning would refer to the quantum fields living in the semiclassical description of the black hole interior. As we explained above, depending on which side of the quantum extremal surface they lie on, these degrees of freedom can be encoded  either  in the Hilbert space  appearing in the central dogma or the Hilbert space living in the radiation region.

\begin{figure}[ht]
\begin{center}
\includegraphics[scale=.35]{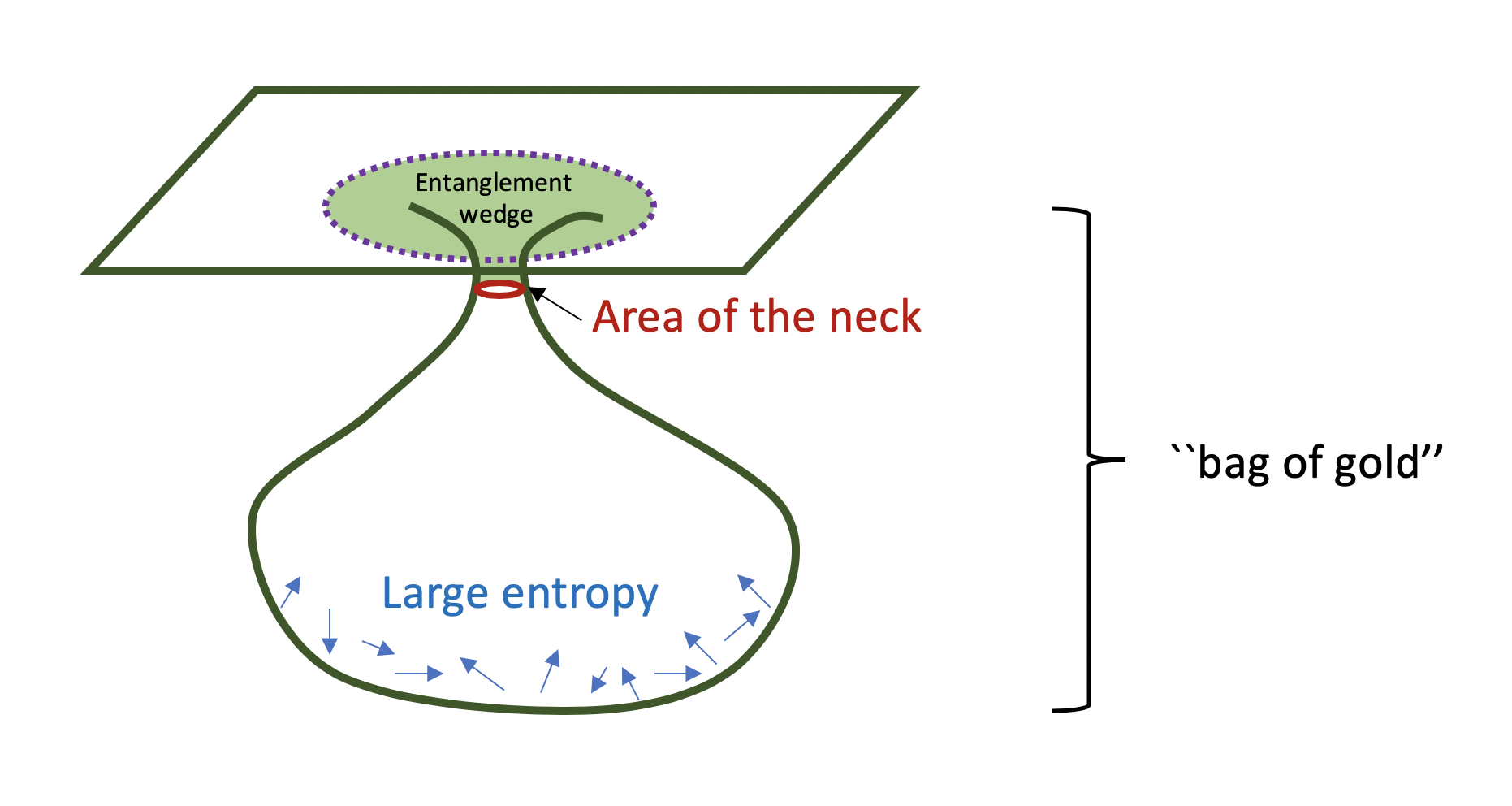}  
\caption{  Wheeler's ``bag of gold'' geometry. It is a spatial geometry that is asymptotically flat and has a narrow neck with a big ``bag'' containing some matter. It obeys the initial value constraints of general relativity. From the point of view of the outside the geometry evolves into a black hole whose area is given by the area of the neck. The entropy inside the bag can be much larger than the area of the neck. Under these circumstances the fine-grained entropy of the exterior is  just   given by the  area of the neck and the entanglement wedge does  not include the interior.  }
\label{BagGold}
\end{center}
\end{figure}

	  This observation also solves an old puzzle with the interpretation of the  Bekenstein-Hawking   area formula that was raised by Wheeler \cite{WheelerBag}. 
 He pointed out that there exist classical geometries which look like a black hole from the outside but that inside can have arbitrarily large entropy, larger than the area of the horizon. He named them ``bags of gold," see figure \ref{BagGold}. The solution to this puzzle is the same. When the entropy in the interior is larger than the area of the neck  the entanglement wedge of the black hole degrees of freedom will only cover a portion of the interior, which does not include that large amount of entropy \cite{WallGold}.   In fact, the geometry of an evaporating black hole after the Page time is a bit like that of the ``bag of gold'' examples. 
 
\section{Replica wormholes}
\la{replicas} 

In this section we would like to give a flavor for the derivation of the 
formula for fine-grained entropy \cite{Lewkowycz:2013nqa,Faulkner:2013ana,Dong:2016hjy,Dong:2017xht,Penington:2019kki,Almheiri:2019qdq}. We focus on the island formula \eqref{island} for the entropy of the Hawking radiation \cite{Penington:2019kki,Almheiri:2019qdq}.

To illustrate the principle, we will consider the case when the black hole has evaporated completely and we will ignore details about the last moments of the black hole evaporation, when the interior disconnects from the exterior. For the purposes of computing the entropy,  the geometry is topologically as shown in figure \ref{BabyUniverse}(b). We want to compute the entropy of the final radiation, assuming that the black hole formed from a pure state.

\begin{figure}
\begin{center}
\includegraphics[scale=1]{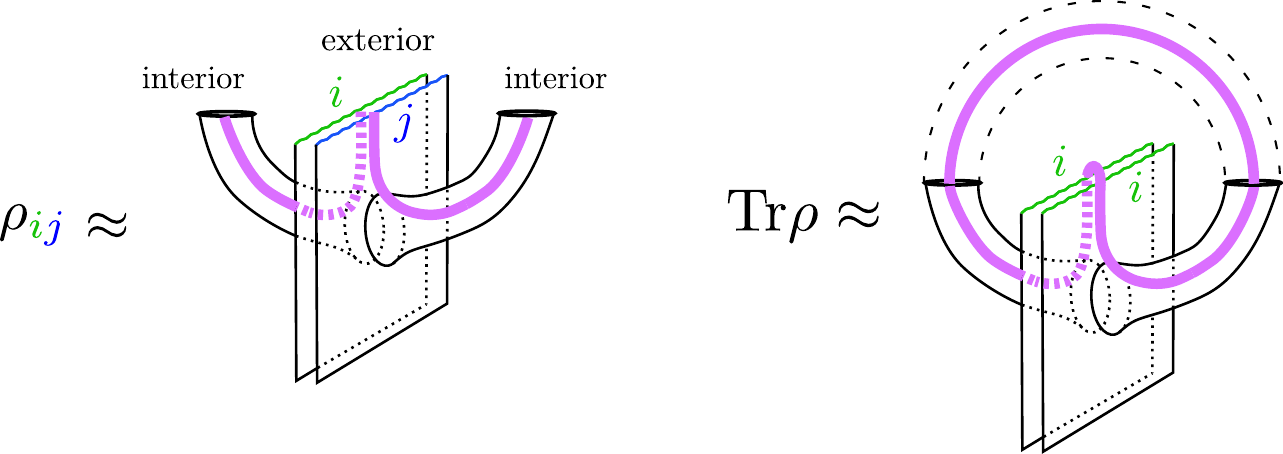}

~~~~~~~~~~~~(a) ~~~~~~~~~~~~~~~~~~~~~~~~~~~~~~~~~~~~~~~~~~~~~~~~~~~~~~~(b)
\end{center}
\caption{\small (a) Path integral representation of the matrix elements $\rho_{ij}$. (b) Path integral representation of $\tr \rho$. Regions with repeated indices are identified in this figure and the figures that follow. The purple line represents entanglement. \label{fig:rhoLorentzianA}}
\end{figure}

  We start from the (unnormalized) initial state $|\Psi\rangle$ $-$ for example a collapsing star $-$   and evolve to the final state using the gravitational path integral which involves the semiclassical geometry of the evaporating black hole. This gives an amplitude $\langle j | \Psi\rangle$ for going from the initial state to a particular final state of radiation, $|j\rangle $ . We can now form a density matrix $\rho = |\Psi\rangle\langle \Psi|$ by computing the bra of the same state via a path integral. Its matrix elements $\rho_{ij} = \langle i |\Psi\rangle\langle \Psi|j\rangle$ are, in principle,  computed by the full gravitational path integral in figure \ref{fig:rhoLorentzianA}a. We have specified the initial and final state on the outside, but we have not been very explicit about what we do in the interior yet, and indeed, this will depend on a choice of $|i\rangle $ and $|j\rangle$.

\begin{figure}
\begin{center}
\includegraphics[scale=1]{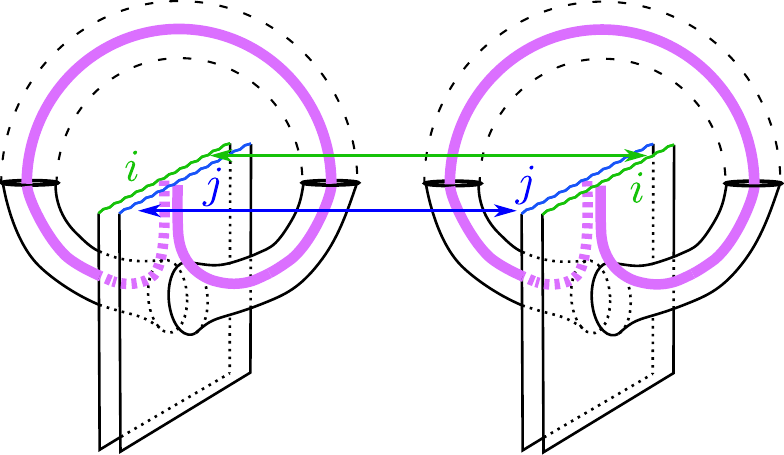}
\end{center}
\caption{\small Hawking saddle in the calculation of $\Tr[\rho^2]$. Note that following the pink line through the identifications $i \leftrightarrow i$ and $j \leftrightarrow j$ produces just one closed loop. Therefore this does not factorize into two copies of $\tr \rho$.\label{fig:rhoLorentzianHawking}}
\end{figure}

\begin{figure}
\begin{center}
\includegraphics[scale=0.95]{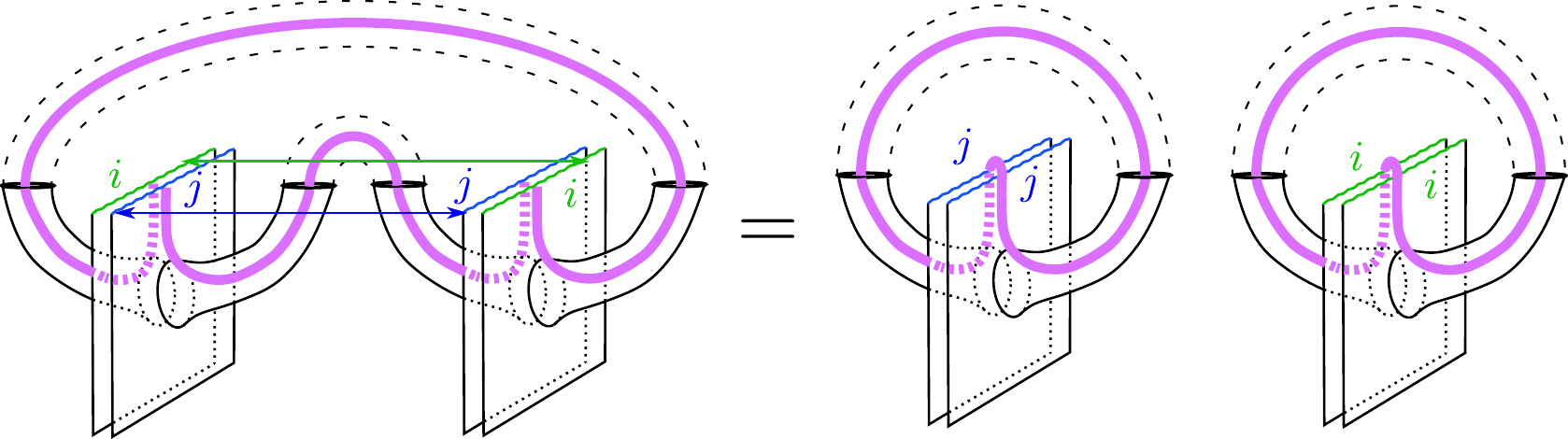}
\end{center}
\caption{\small Replica wormhole saddle in the calculation of $\Tr[\rho^2]$. The black holes are joined in the interior. The second figure is just a rearrangement of the first, showing that $\Tr[\rho^2] = (\Tr[\rho])^2$. \label{fig:rhoLorentzianWormhole}}
\end{figure}

 The trace of the density matrix, 
  \be
\tr \rho = \sum_{i} \langle i|\Psi\rangle\langle \Psi|i\rangle \ , 
  \ee
  is computing by identifying the final state of the bra and the ket and summing over them. This gives
  the geometry in figure \ref{fig:rhoLorentzianA}b. (For those who know, this is really an in-in Schwinger-Keldysh diagram.)
   We want to diagnose whether the final state has zero entropy or not. 
  For that purpose, we compute the so-called ``purity'' of the state, defined as $\tr[\rho^2]$. If $\rho$ is an unnormalized {\it pure state} density matrix then $\tr[\rho^2] = ( \tr[\rho])^2$, while if it has a large entropy we expect $ \tr[\rho^2 ]   \ll ( \tr[\rho])^2 $.

  We can compute $\tr[\rho^2]$ via a path integral argument by connecting the exterior regions as shown in figures \ref{fig:rhoLorentzianHawking} and \ref{fig:rhoLorentzianWormhole}. A key point is that, in gravity, we should typically consider a sum over all possible topologies.\footnote{ This sum is very clearly required in some examples of AdS/CFT to match CFT properties
  \cite{Witten:1998zw}.}   This implies that 
   we should sum over different ways of   connecting the interiors.  Figures \ref{fig:rhoLorentzianHawking} and  \ref{fig:rhoLorentzianWormhole} show two different ways of connecting the interior. The first diagram, 
   figure \ref{fig:rhoLorentzianHawking}, gives the Hawking answer with its large entropy, so that 
  \be \tr[\rho^2 ]|_{\rm Hawking~saddle}  \ll ( \tr[\rho])^2 \la{HSad} \ .
  \ee 
    The second diagram, figure \ref{fig:rhoLorentzianWormhole}, which is called a replica wormhole, gives 
   \be \la{WSad}
   Tr[\rho^2]|_{\rm Wormhole~saddle}  = (\tr[\rho])^2
   \ee
    and therefore has zero entropy. The contribution of the replica wormhole is larger and it therefore dominates over the Hawking saddle \nref{HSad}.      We conclude that the leading order contribution gives the expected answer from unitarity, \nref{WSad}. 

The contribution in figure \ref{fig:rhoLorentzianHawking} is still present and one could worry that it would spoil the agreement. We will not worry about exponentially small contributions,  hoping that this (small) problem will be fixed in the future. 
   
\begin{figure}
\begin{center}
\includegraphics[scale=1]{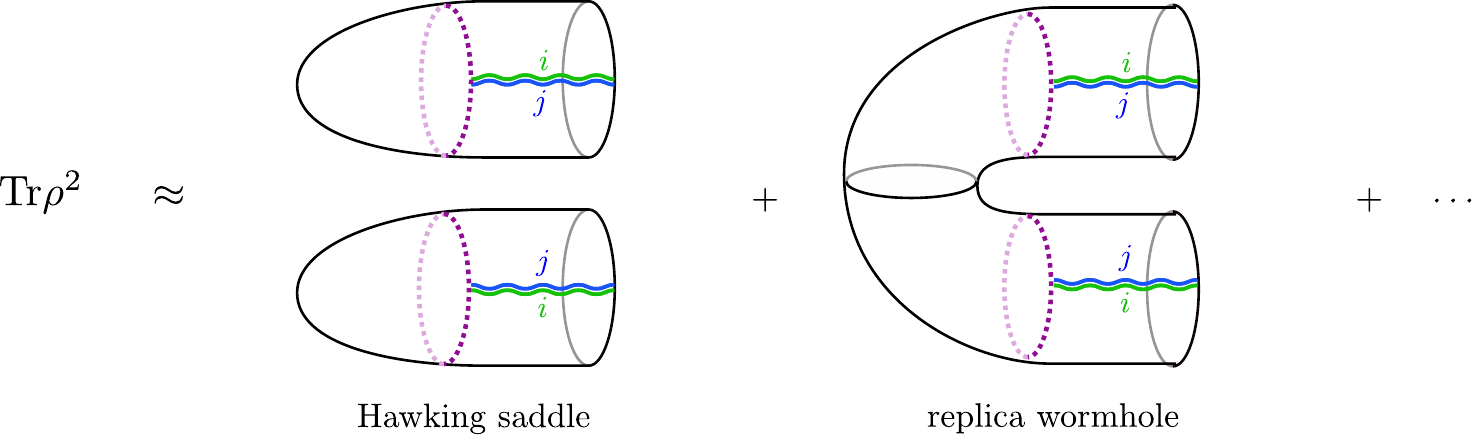}
\end{center}
\caption{\small Euclidean replica wormholes for computing the purity of the radiation outside the cutoff surface. The dots denote other possible topologies, which are generally subdominant. \label{fig:euclidean-wormholes}}
\end{figure}

   This calculation is very similar to the Gibbons-Hawking calculation of the black hole entropy reviewed in section \ref{ss:gibbonshawking}. The Hawking saddle and the replica wormhole saddle in Euclidean signature are drawn in fig. \ref{fig:euclidean-wormholes}. In the Hawking calculation we have two copies of the cigar geometry, while in the replica wormhole the black holes are joined through the interior.  These pictures are schematic because the actual replica wormhole for an evaporating black hole is a complex saddlepoint geometry.
   
   The calculation of the von Neumann entropy is a bit more complicated, but the spirit is the same. We use the replica method. That is, to compute the entropy,  we consider $n$ copies of the system and compute $\tr[\rho^n]$, where $\rho$ is the density matrix of either the black hole or the radiation. We then analytically continue in $n$ and compute the entropy,
\be
S = (1-n \p_n) \log\tr [\rho^n]_{n=1} \ .
\ee
For $n \neq 1$, the black hole interior can be connected in various ways among the $n$ copies. If they are disconnected we get the usual Hawking answer for the entropy, and if they are completely connected we get the answer from the new quantum extremal surface, after continuing to $n \to 1$. The minimum of the two dominates the path integral and gives the unitary Page curve, see \cite{Penington:2019kki,Almheiri:2019qdq} (see also \cite{Marolf:2020xie,Hartman:2020swn}).


\section{Discussion} 
\la{Discussion} 

\subsection{Short summary} 

Let us summarize some of the points we made in the review. 
First we discussed classic results in black hole thermodynamics, including Hawking radiation and black hole entropy. The entropy of the black hole is given by the area of the horizon plus the entropy of the quantum fields outside. 
We discussed how these results inspired a central dogma which says that a black hole from the outside can be described in terms of a quantum system with a number of degrees of freedom set by the entropy. 
Next we discussed a formula for the fine-grained entropy of the black hole which involves finding a surface that minimizes the area plus the entropy of quantum fields outside.
Using this formula,  we computed the entropy for an evaporating black hole and found that it follows the Page curve. Then we discussed how to compute the entropy of radiation. The gravitational fine-grained entropy formula tells us that we should include the black hole interior and it gives a result that follows the Page curve, too. 
These results suggest that the black hole degrees of freedom describe a portion of the interior, the region inside the entanglement wedge. 
Finally we discussed how replica wormholes explain why the interior should be included in the computation of the entropy of radiation.  

\subsection{Comments and problems for the future } 

It is important to point out the following important feature of the gravitational entropy formulas, both the coarse-grained and the fine-grained one. Both formulas involve a geometric piece, the area term, which does not obviously come from taking a trace over some explicit microstates. The interpretation of these quantities as arising from sums over microstates is an assumption, a part of the ``central dogma," which is the simplest way to explain the emergence of black hole thermodynamics, and  has strong evidence from string theory.

For this reason, the success in reproducing the Page curve does not translate into a formula for individual matrix elements of the density matrix.  The geometry is giving us the correct entropy, which involves a trace of (a function of) the density matrix. 
Similarly we do not presently know how to extract 
  individual matrix elements of the black hole S-matrix, which describes individual transition amplitudes for each microstate. Therefore the current discussion leaves an important problem unresolved. Namely, how do we compute individual matrix elements of the S-matrix, or $\rho$, directly from the gravity description (without using a holographic duality)? In other words, we have discussed how to compute the entropy of Hawking radiation, but not how to compute its precise quantum  state. This is an important aspect of the black hole information problem, since one way of stating the problem is: Why do different initial states lead to the same final state? In this description the different initial states correspond to different interiors. In gravity, we find that the final state for radiation also includes the interior.  
  The idea is that very complex computations in the radiation can create wormholes that reach into that interior and pull out the information stored there \cite{Penington:2019kki}, 
  see also \cite{Maldacena:2013xja,Gao:2016bin}.

The present derivations for the 
  gravitational fine-grained entropy formulas discussed in this paper rely on the Euclidean path integral. It is not clear how this is defined precisely in gravity. For example, which saddle points should we include? What is the precise integration contour? It is possible that some theories of gravity include replica wormhole saddles and black holes evaporate unitarily, while in other theories of gravity they do not contribute to the path integral, the central dogma fails, and Hawking's picture is accurate. (We suspect that the latter would not be fully consistent theories.)

Another aspect of the formulas which is not yet fully understood is the imaginary cutoff surface, beyond which we treated spacetime as fixed. This is an important element in the derivation of the formula \eqref{island}  as discussed in section \ref{replicas}.   
 A more complete understanding will require allowing gravity to fluctuate everywhere throughout spacetime. For example, we do not know whether the central dogma applies when the cutoff is at a finite distance from the black hole, or precisely how far we should go in order to apply these formulas. The case that is best understood is when this cutoff is at the boundary of an AdS space.   On the other hand, the imaginary cutoff surface is not as drastic as it sounds because the same procedure is required to make sense of the ordinary Gibbons-Hawking entropy in asymptotically flat spacetime.

Note that when we discussed the radiation, we had  two quantum states in mind. First we had the semiclassical state, the state of radiation that appears when we use the semiclassical geometry of the evaporating black hole. Then we had the exact quantum state of radiation. This is the state that would be produced by the exact and complete theory of quantum gravity. Presumably, to obtain this state we will need to sum over all geometries, including non-perturbative corrections. This is something that we do not know how to do in any theory of gravity complicated enough to contain quantum fields describing Hawking radiation. (See however \cite{Saad:2019lba,Penington:2019kki} for some toy models.) The magic of the gravitational fine-grained entropy formula is that it gives us the entropy of the exact state in terms of quantities that can be computed using the semiclassical state. One could ask, if you are an observer in the radiation region, which of these two states should you use? If you do simple observations, the semiclassical state is good enough. But if you consider very complex observables, then you need to use the exact quantum state. One way to understand this is that very complex operations on the radiation weave their own spacetime, and this spacetime can develop a connection to the black hole interior. See \cite{Susskind:2018pmk} for more discussion.

  This review has focused on novel physics in the presence of black hole event horizons. In our universe, we also have a cosmological event horizon due to accelerated expansion. This horizon is similar to a black hole horizon in that it has an associated Gibbons-Hawking entropy and it Hawking radiates at a characteristic temperature 
\cite{Figari:1975km,Gibbons:1977mu}. 
However, it is unclear whether we should think of the cosmological horizon as a quantum system in the sense of the central dogma for black holes. Applying the ideas developed in the previous sections to cosmology may shed light on the nature of these horizons and the quantum nature of cosmological spacetimes.

   There is a variant of the black hole information problem where one perturbs the black hole and then looks at the response at a very late time in the future \cite{Maldacena:2001kr}.  For recent progress in that front see \cite{Saad:2018bqo,Saad:2019lba,Saad:2019pqd}.

      Wormholes similar to the ones discussed here were considered in the context of theories with random couplings \cite{Coleman:1988cy,Giddings:1988cx,Polchinski:1994zs}. Recently, random couplings played an important role in the solution of a simple two dimensional gravity theory \cite{Saad:2018bqo,Saad:2019lba}. 
      We do not know to what extent random couplings are important for the issues we discussed in this review. See also \cite{Marolf:2020xie}.

 We should emphasize one point. In this review,  we have presented results that can be understood purely in terms of gravity as an effective theory. However, string theory and holographic dualities played an  instrumental role in inspiring and checking these results. They provided concrete examples where these ideas were tested and developed,  before they were applied to the study of black holes in general. 
 Also, as we explained in the beginning, we have not followed a  historical route and we have not reviewed ideas that have led to the present status of understanding.    
   
  Finally, we should finish with a bit of a cautionary tale. Black holes are very confusing and many researchers who have written papers on them have gotten some things right and some wrong.  What we have discussed in this review is an {\it interpretation} of some geometric gravity computations. We interpreted them in terms of entropies of quantum systems. It could well be that our interpretation will have to be revised in the future, but we have strived to be conservative and to present something that is likely to stand the test of time.

A goal of quantum gravity is to understand what spacetime is made of. The fine-grained entropy formula is giving us very valuable  information on how the fundamental quantum degrees of freedom are building the spacetime geometry. These studies have involved the merger and ringdown of several different fields of physics over the last few decades: high energy theory, gravitation, quantum information, condensed matter theory, etc.,  creating connections beyond their horizons. This has not only provided exciting insights into the quantum mechanics of black holes, but also turned black holes into a light that illuminates many questions of these other fields. Black holes have become a veritable source of information!

\vspace{1cm}
\textbf{Acknowledgments} We are grateful to R. Bousso, A. Levine, A. Lewkowycz, R. Mahajan, S. Shenker, D. Stanford, A. Strominger, L. Susskind, A. Wall   and 
 Z. Yang 
for helpful discussions on these topics. We also thank G. Musser for comments on a draft of this review.
 
A.A. is supported by funds from the Ministry of Presidential Affairs, UAE. The work of ES is supported by the Simons Foundation as part of the Simons Collaboration on the Nonperturbative Bootstrap. The work of TH and AT is supported by DOE grant DE-SC0020397.
J.M. is supported in part by U.S. Department of Energy grant DE-SC0009988 and by the Simons Foundation grant 385600.

  \appendix
  

\section{Comments on the AMPS paradox } 

In \cite{Almheiri:2012rt}  a  problem or paradox  was found,  and a proposal was made for its resolution.   Briefly stated, the paradox was the impossible quantum state appearing after the Page time, where the newly outgoing Hawking quantum needs to be maximally entangled with two seemingly separate systems: its interior partner and the early Hawking radiation. The proposed resolution was to declare the former entanglement broken, forming a ``firewall'' at the horizon.
A related problem was discussed in \cite{Marolf:2012xe}. 

The paradox involved the central dogma plus one extra   implicit assumption. 
The extra assumption is that the black hole interior can also be described by the {\it same} 
degrees of freedom that describe the black hole from the outside, the degrees of freedom that appear in the central dogma.
 We have not made this assumption in this review. 

According to this review,  the  paradox is resolved by dropping the assumption that the interior is also described by the same degrees of freedom that describe it as viewed from outside.     Instead,  we assume that only a portion of the interior is described by the black hole degrees of freedom appearing in the central dogma $-$   only the portion in the entanglement wedge, see figure \ref{EWfig}(b).  This leaves the interior as part of the radiation, and the resolution of the apparently impossible quantum state is that the interior partner is identified with part of the early radiation that the new Hawking quantum is entangled with.
This is different than the resolution proposed in AMPS. With this resolution, the horizon is smooth.

\section{Glossary}  

{\bf Causal diamond}: The spacetime region that can be determined by evolution (into the future or the past) of initial data on any spatial region.  See figure \ref{Diamond}. \\
\\
{\bf Central dogma}: A black hole -- as viewed from the outside -- is simply a quantum system with a number of degrees of freedom equal to Area$/4G_N$. Being a quantum system, it evolves unitarily under time evolution. See section \ref{central}. \\
\\
{\bf Fine-grained entropy}: Also called the von Neumann entropy or quantum entropy. Given a density matrix $\rho$, the fine-grained entropy is given as $S = -Tr[\rho \log \rho]$. See section \ref{finecoarse}.\\
\\
{\bf Coarse-grained entropy}: Given a density matrix $\rho$ for our system, we measure a subset of simple observables $A_i$ and consider all $\tilde{\rho}$ consistent with the outcome of our measurements, $Tr[\tilde{\rho} A_i] = Tr[\rho A_i]$. We then maximize the entropy $S(\tilde{\rho}) = -Tr[\tilde{\rho} \log \tilde{\rho}]$ over all possible choices of $\tilde{\rho}$. See section \ref{finecoarse}. \\
\\
{\bf Semiclassical entropy}: The fine-grained entropy of matter and gravitons on a fixed background geometry. See section \ref{semiclassical}. \\
\\
{\bf Generalized entropy}: The sum of an area term and the semi-classical entropy.  See \eqref{sgendef}. When evaluated at an event horizon soon after it forms, for example in \eqref{sgen}, the generalized entropy is coarse grained. When evaluated at the extremum, as in \eqref{RT} or \eqref{island}, the generalized entropy is fine grained. \\
\\
{\bf Gravitational fine-grained entropy}: Entropy given by the formulas \nref{RT} and \nref{island}. They give the fine-grained entropy through a formula that involves a geometric part, the area term, and the semiclassical entropy of the quantum fields.  \\
\\
{\bf Page curve}: Consider a spacetime with a black hole formed by the collapse of a pure state. Surround the black hole by an imaginary sphere whose radius is a few Schwarzschild radii. The Page curve is a plot of the fine-grained entropy outside of this imaginary sphere, where we subtract the contribution of the vacuum. Since the black hole Hawking radiates and the Hawking quanta enter this faraway region, this computes the fine-grained entropy of Hawking radiation as a function of time. Notice that the regions inside and outside the imaginary sphere are open systems. The curve begins at zero when no Hawking quanta have entered the exterior region, and ends at zero when the black hole has completely evaporated and all of the Hawking quanta are in the exterior region. The ``Page time" corresponds to the turnover point of the curve. See figure \ref{HawkingPageCurves}.\\
\\
{\bf Quantum extremal surface}: The surface $X$ that results from extremizing (and if necessary minimizing) the generalized entropy as in \eqref{RT}. This same surface appears as a boundary of the island region in \eqref{island}. \\
\\
{\bf Island}: Any  disconnected codimension-one regions found by the extremization procedure \eqref{island}. Its boundary is the quantum extremal surface. The causal diamond of an island region is a part of the entanglement wedge of the radiation.\\
\\
{\bf Entanglement wedge}: For a given system (in our case either the radiation or the black hole), the entanglement wedge is a region of the semiclassical spacetime that is described by the system.  It is defined at a moment in time and has nontrivial time dependence. Notice that language is not a good guide: the transition in the Page curve from increasing entropy to decreasing entropy corresponds to when most of the interior of the black hole becomes described by the radiation, i.e. the entanglement wedge of the black hole degrees of freedom does not include most of the black hole interior. See section \ref{wedge} and figure \ref{EWfig}. \\
\\
{\bf Replica trick}: A mathematical technique used to compute $-Tr[\rho \log \rho]$ in a situation where we do not have direct access to the matrix $\rho_{ij}$. See section \ref{replicas}.\\
\\

  \end{spacing}
\small
\bibliographystyle{ourbst}
 \bibliography{ReviewDraft}

\end{document}